\documentclass[a4paper,12pt]{article}
\usepackage{amsmath}
\usepackage{amssymb}
\usepackage{amsmath,amsthm,amssymb,amscd}
\usepackage{latexsym}
\usepackage{indentfirst}
\usepackage{subfigure}
\usepackage{graphicx}
\usepackage[top=1in,bottom=1in,left=1.25in,right=1.25in]{geometry}

\makeatother
\date{}
\theoremstyle{plain}

\begin{document}

\bibliographystyle{unsrt}
\bibliographystyle{plain}

\title{
  \bf Properties of tug-of-war model for cargo transport by molecular motors
}

\author{{Yunxin Zhang}\thanks{
School of Mathematical Sciences, Fudan University,  Shanghai 200433,
China}
\thanks{Shanghai Key Laboratory for Contemporary Applied Mathematics, Fudan
University}
\thanks{Centre for Computational Systems Biology, Fudan University   (E-Mail: xyz@fudan.edu.cn)}
}

\maketitle \baselineskip=6mm

\baselineskip=22pt

\begin{abstract}
    Molecular motors are essential components for the
biophysical functions of the cell. Current quantitative
understanding of how multiple motors move along a single track is
not complete, even though models and theories for a single motor
mechanochemistry abound. Recently, M.J.I. M\"{u}ller {\em et al.}
have developed a tug-of-war model to describe the bidirectional
movement of the cargo (PNAS(2008) 105(12) P4609-4614). %Through Monte
%Carlo simulations,
They found that the tug-of-war model exhibits several qualitative
different motility regimes, which depend on the precise value of
single motor parameters, and they suggested the sensitivity can be
used by a cell to regulate its cargo traffic.  In the present paper,
we will carry out a further detailed theoretical analysis of the
tug-of-war model. All the stable, i.e., biophysically observable,
steady states and their stability domains can be obtained. Depending
on values of the several parameters, tug-of-war model exhibits
either uni-, bi- or tristability. In large motor number case, the
steady state movement of the cargo, which is transported by two
molecular motor species, is determined by the initial numbers of the
motors which bound to the track. For small motor number case, the
movement of cargo may jump from one of the stable steady state to
another.

%Based on the separating boundary of the different stable states and
%the initial numbers of the different motor species that are bound to
%the track, the steady state of the cargo movement can be predicted,
%and consequently the steady state velocity can be
%obtained. %It is found that, the velocity, even the direction, of the
%cargo movement change with the initial numbers of the motors which
%are bound to the track and several other parameters.

 \vspace{2em} \noindent \textit{PACS}: 87.16.Nn, 87.16.A-,
82.39.-k, 05.40.Jc

\vspace{2em} \noindent \textit{Keywords}: Tug-of-war, molecular
motors, intracellular transport
\end{abstract}

\section{Introduction}
Molecular motors, including biological motor proteins such as
kinesin \cite{David2007,Carter2005,Taniguchi2005,Zhang20081}, dynein
\cite{Vale2003,Sakakibara1999}, mysion
\cite{Hooft2007,Christof2006,Katsuyuki2007} and $F_0F_1$-ATP
synthase \cite{Noji1997}, are mechanochemical force generators which
convert chemical or biochemical energy in the form of chemical
potential into mechanical work in thermal environment
\cite{Howard2001}.  The mechanochemical process is accomplished by
individual macromolecules, immersed in an aqueous solution with the
chemical potential, moving along a linear track.  Many biological
motor proteins move processively. For example, myosin slides along
an actin filament, kinesin and dynein along microtubule (MT). All of
them are adenosine triphosphate (ATP)-driven ``directional walking
machines'' (\cite{Wang1998,Masayoshi2002}): Kinesin moves towards
the plus end of the MT and dynein towards the minus end. In
comparison with the macroscopic engines driven by Carnot cycles,
molecular motors have a high energy efficiency at about 50\%, while
the energy efficiency of a car is about 15\%-20\%
\cite{Vale2003,Wang2005,Zhang2008}. Furthermore, the velocities of
molecular motors are also fast with mean velocity be at about
several hundreds nanometers per second \cite{voboda1994}. However,
the most significant difference between the molecular motors and the
macroscopic engines is that the former are moving in a thermal noise
dominated environment \cite{Reimann2002}. So the movement of the
molecular motors should be described stochastically, rather than
determinately.  Being able to convert and harvest energy with high
efficiency on a mesoscopic scale makes molecular motors an exciting
area of scientific research with potentially great innovative
applications for energy production.

\begin{figure}
  \centering
  \includegraphics[width=300pt]{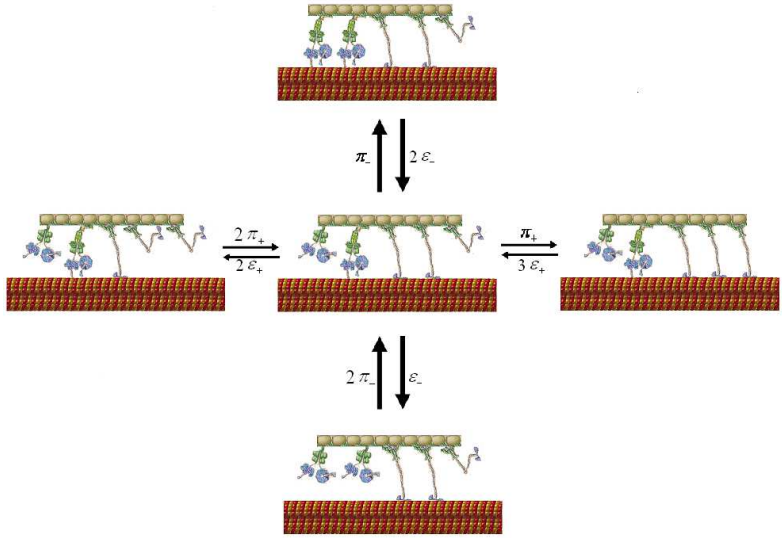}\\
  \caption{Schematic depiction of tug-of-war model: A cargo with $N_+=3$ plus motors (Kinesin) and $N_-=2$ motors (Dynein) is pulled by a fluctuating
  number of motors bound to the microtubule.
  }\label{Fig1}
\end{figure}

Great progress has been made in recent years in modeling the
movement of molecular motors, including the mean field methods
\cite{Wang2008,Qian2000,Howard2001}, the Langevin stochastic dynamic
methods \cite{Reimann2001,Astumian1997} and discrete stochastic
methods \cite{Astumian2005,Liepelt2007,Fisher2001,
Derrida1997,Qian1997}. However, %one aspect of the motor protein
%movement that has never been fully studied:  When there are multiple
%processive motors attached to a single track, there is an issue of
%``traffic jam''!
the existing models for a single molecular motor are not sufficient
in predicting the recent experimental results: It is found that
bidirectional motion of the cargo, which is carried by motor
proteins, exhibits different patterns in different stages of
embryonic development(\cite{Gross2002}). Following these recent
experimental results (\cite{Deacon2003,Welte2005,Smith2004}),
Lipowsky and his coworkers have developed the tug-of-war model for
describing the movement of the cargo carried by processive motors,
such as kinesin and dynein
(\cite{Lipowsky2008,Beeg2008,Lipowsky2005,Lipowsky20081,Lipowsky20082}).
In their model, the experimentally known single motor properties are
taken into account, so it is consistent with almost all experimental
observations and can make quantitative predictions for bidirectional
transport of the cargo. Since cargo movement carried by a single
motor protein via an elastic tether has been extensively studied in
the past \cite{Chen2002, Elston2000}, the focus of tug-of-war model
is not on the detailed movement of cargo carried by a single motor
{\em per se}, rather it concerns with the competition and
cooperation of multiple motors on a single track (see the schematic
depiction in Fig. \ref{Fig1}).

%The tug-of-war model has been studied in this, and that.  But no
%comprehensive mathematical analysis of the nonlinear mathematical
%equations has be carried out.  Most previous results rely on Monte
%Carlo simulation.
In the present paper, we will give a further comprehensive
mathematical analysis of tug-of-war model. Through detailed
analysis, we find that the steady state movement of cargo is
determined by the initial numbers of the two motor species which bound to the track of movement.
%Given an initial
%numbers of plus-end and minus-end motors (e.g., kinesin and dynein)
%which attached to the track (e.g., MT), the steady state of the
%movement of the cargo can be calculated theoretically, and thus the
%average velocity of the movement can be obtained.
Biophysically, the steady state is the only state that can be
observed experimentally. At the same time, Monte Carlo simulations
indicate the transition time from the initial state to the steady
state is very short (see Figs. \ref{Fig7}, \ref{Fig8}).
Theoretically, the movement of the cargo has at most three stable
steady states. If there exists two or three stable steady states,
then many parameters of plus and minus motors have at least one
critical point. The movement of cargo would change from one stable
steady state to
another if one of the parameters jumps from one side of its critical point to another side. %At steady
%state, the numbers $n_+$ and $n_-$ of the plus and minus motors
%which bound to the track, oscillate around a theoretical stable
%steady state, which can be regarded as the stable steady state with
%numbers of the plus and minus motors, which attached to the cargo,
%tend to infinite.
In the following, we firstly introduce the tug-of-war model, and
then give the detailed discussion gradually.

\section{ The tug-of-war model}
The tug-of-war model is developed by Reinhard Lipowsky's study group
(\cite{Lipowsky2008,Beeg2008,Lipowsky2005,Lipowsky20081,Lipowsky20082})
to study the  bidirectional transport of the cargo, in which the
cargo is attached with $N_+$ plus and $N_-$ minus motors.
Particularly, if $N_+=0$ or $N_-=0$, it recovers the usual model for
cooperate transport by a single motor species (\cite{Lipowsky2005}
\cite{Frank1995}). In this model, each motor species is
characterized by six parameters, which can be measured in single
molecular experiments (see Tab. \ref{table1}): (i) stall force $F_S$
(pN) (ii) detachment force $F_d$ (pN) (iii) unbinding rate
$\epsilon_0$ ($s^{-1}$) (iv) binding rate $\pi_0$ ($s^{-1}$) (v)
forward velocity $v_F$ ($\mu$m/s) and (vi) superstall velocity
amplitude $v_B$ (nm/s). The motors bind to or unbind from a MT in a
stochastic fashion, so that the cargo is pulled by $n_+\le N_+$ plus
and $n_-\le N_-$ minus motors, where $n_+$ and $n_-$ fluctuate with
time (see Fig. \ref{Fig1}).
\begin{table}
  \centering
    \begin{tabular}{lccc}
    \hline
      Parameter&Symbol&Kinesin 1 & Dynein  \\
    \hline \hline
     Stall force &$F_s$&6pN&1.1pN  \\
     Detachment force &$F_d$&3pN&0.75pN  \\
     Unbinding rate &$\epsilon_0$&1$s^{-1}$&0.27$s^{-1}$  \\
     Binding rate &$\pi_0$&5$s^{-1}$&1.6$s^{-1}$  \\
    Forward velocity &$v_F$&1$\mu$m/s&0.65$\mu$m/s  \\
     superstall velocity amplitude &$v_B$&6nm/s&72nm/s  \\
    \hline
  \end{tabular}
  \caption{Single-motor parameters for kinesin 1 and cytoplasmic dynein (\cite{Lipowsky2008} and references therein).}\label{table1}
\end{table}

In tug-of-war model, it is assumed that, at every time $t$, the
state of cargo with $N_+$ plus and $N_-$ minus motors firmly
attached to it is fully characterized by numbers $n_+$ and $n_-$ of
plus and minus motors that are bound to the MT. The state of cargo
changes when a plus or a minus motor binds or unbinds to/from the MT
(see Fig. \ref{Fig1}). %These changes can be described by a master equation for
%the probability distribution $p(n_+,n_-,t)$ of $n_+$ bound plus and
%$n_-$ bound minus motors at time $t$:
The probability $p(n_+,n_-,t)$ to have $n_+$ plus and $n_-$ minus
bound motors at time $t$ can be described by the following Master
equation:
\begin{equation}\label{eq4}
\begin{aligned}
\frac{dp(n_+,n_-,t)}{dt}=&[N_+-(n_+-1)]\pi_+p(n_+-1,n_-,t)\cr
&+(n_++1)\epsilon_+(n_++1,n_-)p(n_++1,n_-,t)\cr
&+[N_--(n_--1)]\pi_-p(n_+,n_--1,t)\cr
&+(n_-+1)\epsilon_+(n_+,n_-+1)p(n_+,n_-+1,t)\cr
&-[(N_+-n_+)\pi_++n_+\epsilon_+(n_+,n_-)\cr
&+(N_--n_-)\pi_-+n_-\epsilon_-(n_+,n_-)]p(n_+,n_-,t)\cr &\quad 1\le
n_+\le N_+-1\quad\textrm{and}\quad 1\le n_-\le N_--1
\end{aligned}
\end{equation}
where $\pi_+ (\pi_-)$ is the binding rate of a single plus (minus)
motor to the MT, which depends only weakly on the load
(\cite{Lipowsky2005}) and therefore is taken equal to zero-load
binding rate $\pi_{0+} (\pi_{0-})$. $\epsilon_+ (\epsilon_-)$ is the
unbinding rate of a single plus (minus) motor from the MT, which
increases exponentially with the applied force $F$:
\begin{equation}\label{eq2}
\epsilon_{\pm}(F)=\epsilon_{0\pm}\exp(|F|/F_{d\pm})
\end{equation}
as measured for kinesin \cite{Block1994}, where $F_d$ is the
detachment force. The governing equations for $n_+=0, N_+$ or
$n_-=0, N_-$ are similar as (\ref{eq4}) except
$\pi_+(N_+,n_-)=\pi_-(n_+,N_-)=0$ and
$\epsilon_+(0,n_-)=\epsilon_-(n_+,0)=0$.

Under the assumptions that the motors act independently and feel
each other only due to two effects: (i) opposing motors act as load,
and (ii) identical motors share this load, Lipowsky and coworkers
gave the following relation (see \cite{Lipowsky20081})
\begin{equation}\label{eq5}
n_+F_+=-n_-F_-\equiv F_c
\end{equation}
where $F_+ (-F_-)$ is the load felt by each plus (minus) motor. Eqs.
(\ref{eq2}) (\ref{eq5}) imply
\begin{equation}\label{eq6}
\epsilon_\pm(n_+,n_-)=\epsilon_{0\pm}\exp[|F_c|/n_\pm F_{d\pm}]
\end{equation}
Here, the cargo force $F_c$ is determined by the condition that plus
motors, which experience the force $F_c/n_+$, and minus motors,
which experience the force $-F_c/n_-$, move with the same velocity
$v_c$, which is the cargo velocity:
\begin{equation}\label{eq7}
v_c(n_+,n_-)=v_+(F_c/n_+)=-v_-(-F_c/n_+)
\end{equation}
The same as in \cite{Lipowsky2008}, the following piecewise linear
force-velocity relation of a single motor is used in this paper:
\begin{equation}\label{eq1}
v(F)=\left\{
\begin{aligned}
&v_F(1-F/F_s)\qquad\textrm{for}\qquad F\le F_s\cr
&v_B(1-F/F_s)\qquad\textrm{for}\qquad F\ge F_s
\end{aligned}
\right.
\end{equation}
where $v_B$ is the absolute value of the superstall velocity
amplitude, $v_F$ is the zero-load forward velocity, $F_s$ is the
stall force.

\section{The velocity of cargo and unbinding rates of motors}
For the convenience of analysis in the following sections, we give
the formulations of velocity of cargo and unbinding rates of plus
and minus motors in this section.

\noindent{\bf (I) }In case of \lq\lq stronger plus motors\rq\rq,
i.e. $n_+F_{s+}>n_-F_{s-}$, Eqs. (\ref{eq7}) (\ref{eq1}) lead to the
cargo force and velocity:
\begin{equation}\label{eq8}
\begin{aligned}
%F_c(n_+,n_-)&=\frac{n_+n_-F_{s+}F_{s-}(v_{F+}+v_{B-})}{n_+F_{s+}v_{B-}+n_-F_{s-}v_{F+}}\cr
F_c(n_+,n_-)&=\frac{v_{F+}+v_{B-}}{v_{F+}/n_+F_{s+}+v_{B-}/n_-F_{s-}}\cr
v_c(n_+,n_-)&=\frac{n_+F_{s+}-n_-F_{s-}}{n_+F_{s+}/v_{F+}+n_-F_{s-}/v_{B-}}
\end{aligned}
\end{equation}
Using Eqs. (\ref{eq6}) (\ref{eq8}), the unbinding rates of plus and
minus motors are:
\begin{equation}\label{eq9}
\begin{aligned}
\epsilon_{\pm}(n_+,n_-)=&\epsilon_{0\pm}\exp\left(\frac{n_\mp
F_{s+}F_{s-}(v_{F+}+v_{B-})}{(n_+F_{s+}v_{B-}+n_-F_{s-}v_{F+})F_{d\pm}}\right)\cr
=&:\epsilon_{0\pm}\exp\left(\frac{n_\mp
}{(an_++bn_-)F_{d\pm}}\right)
\end{aligned}
\end{equation}
where
\begin{equation}\label{eq10}
a=\frac{v_{B-}}{F_{s-}(v_{F+}+v_{B-})}\qquad
b=\frac{v_{F+}}{F_{s+}(v_{F+}+v_{B-})}
\end{equation}
Let $x=n_+/N_+$, $y=n_-/N_-$ and $c=N_+/N_-$, then
\begin{equation}\label{eq101}
\begin{aligned}
\epsilon_{+}(x,y)=&\epsilon_{0+}\exp\left(\frac{y
}{(acx+by)F_{d+}}\right)\cr
\epsilon_{-}(x,y)=&\epsilon_{0-}\exp\left(\frac{cx
}{(acx+by)F_{d-}}\right)
\end{aligned}
\end{equation}

\noindent{\bf (II) }In case of \lq\lq stronger minus motors\rq\rq,
i.e. $n_+F_{s+}<n_-F_{s-}$, the cargo force and velocity are:
\begin{equation}\label{eq102}
\begin{aligned}
F_c(n_+,n_-)&=-\frac{v_{B+}+v_{F-}}{v_{B+}/n_+F_{s+}+v_{F-}/n_-F_{s-}}\cr
v_c(n_+,n_-)&=-\frac{n_-F_{s-}-n_+F_{s+}}{n_+F_{s+}/v_{B+}+n_-F_{s-}/v_{F-}}\cr
&=-\frac{yF_{s-}-xcF_{s+}}{xcF_{s+}/v_{B+}+yF_{s-}/v_{F-}}
\end{aligned}
\end{equation}
Similar as in {\bf (I)}, the unbinding rates of plus and minus
motors are
\begin{equation}\label{eq103}
\begin{aligned}
\epsilon_{+}(x,y)=&\epsilon_{0+}\exp\left(\frac{y
}{(\bar{a}cx+\bar{b}y)F_{d+}}\right)\cr
\epsilon_{-}(x,y)=&\epsilon_{0-}\exp\left(\frac{cx
}{(\bar{a}cx+\bar{b}y)F_{d-}}\right)
\end{aligned}
\end{equation}
in which
\begin{equation}\label{eq104}
\bar{a}=\frac{v_{F-}}{F_{s-}(v_{B+}+v_{F-})}\qquad
\bar{b}=\frac{v_{B+}}{F_{s+}(v_{B+}+v_{F-})}
\end{equation}

The splitting boundary of case {\bf (I)} and case  {\bf (II)} is
$n_+F_{s+}=n_-F_{s-}$, i.e. $y=xcF_{s+}/F_{s-}$.

\noindent{\bf (III) } If an external force $F_{ext}$ is present,
here $F_{ext}$ is taken to be positive if it points into the minus
direction, then the force balance (\ref{eq5}) becomes
$$
n_+F_+=-n_-F_+F_{ext}
$$
In case of $n_+F_{s+}-F_{ext}>n_-F_{s-}$, carrying through the same
calculation as for the case without external force leads to the
velocity of cargo
\begin{equation}\label{eq105}
\begin{aligned}
v_c(n_+,n_-)&=\frac{n_+F_{s+}-n_-F_{s-}-F_{ext}}{n_+F_{s+}/v_{F+}+n_-F_{s-}/v_{B-}}
\end{aligned}
\end{equation}
The corresponding unbinding rates of the plus and minus motors are
\begin{equation}\label{eq106}
\begin{aligned}
\epsilon_{+}(x,y)=&\epsilon_{0+}\exp\left(\frac{y
+aF_{ext}/N_-}{({a}cx+{b}y)F_{d+}}\right)\cr
\epsilon_{-}(x,y)=&\epsilon_{0-}\exp\left(\frac{cx
-bF_{ext}/N_-}{({a}cx+{b}y)F_{d-}}\right)
\end{aligned}
\end{equation}

\noindent{\bf (IV) } If an external force $F_{ext}$ is present and
$n_+F_{s+}-F_{ext}<n_-F_{s-}$, then the velocity of cargo is
\begin{equation}\label{eq107}
\begin{aligned}
v_c(n_+,n_-)&=\frac{n_+F_{s+}-n_-F_{s-}-F_{ext}}{n_+F_{s+}/v_{B+}+n_-F_{s-}/v_{F-}}
\end{aligned}
\end{equation}
and the unbinding rates of plus and minus motors are
\begin{equation}\label{eq108}
\begin{aligned}
\epsilon_{+}(x,y)=&\epsilon_{0+}\exp\left(\frac{y
+\bar{a}F_{ext}/N_-}{(\bar{a}cx+\bar{b}y)F_{d+}}\right)\cr
\epsilon_{-}(x,y)=&\epsilon_{0-}\exp\left(\frac{cx
-\bar{b}F_{ext}/N_-}{(\bar{a}cx+\bar{b}y)F_{d-}}\right)
\end{aligned}
\end{equation}

Similarly, the splitting boundary of case {\bf (III)} and case  {\bf
(IV)} is $n_+F_{s+}=n_-F_{s-}+F_{ext}$, i.e.
$y=xcF_{s+}/F_{s-}-F_{ext}/N_-F_{s-}$.

\noindent{\bf (V) } More generally, if there exists an external
force $F_{ext}$ and the friction coefficient of cargo is $\gamma$,
then in the case of $n_+F_{s+}-F_{ext}>n_-F_{s-}$, the velocity of
the cargo is
\begin{equation}\label{eq109}
\begin{aligned}
v_c(n_+,n_-)&=\frac{n_+F_{s+}-n_-F_{s-}-F_{ext}}{n_+F_{s+}/v_{F+}+n_-F_{s-}/v_{B-}+\gamma}
\end{aligned}
\end{equation}
and the unbinding rates of plus and minus motors are
\begin{equation}\label{eq110}
\begin{aligned}
\epsilon_{+}(x,y)=&\epsilon_{0+}\exp\left(\frac{y +a(F_{ext}+\gamma
v_c)/N_-}{({a}cx+{b}y)F_{d+}}\right)\cr
\epsilon_{-}(x,y)=&\epsilon_{0-}\exp\left(\frac{cx -b(F_{ext}+\gamma
v_c)/N_-}{({a}cx+{b}y)F_{d-}}\right)
\end{aligned}
\end{equation}
On the other hand, if $n_+F_{s+}-F_{ext}<n_-F_{s-}$, then the
velocity of cargo is
\begin{equation}\label{eq111}
\begin{aligned}
v_c(n_+,n_-)&=\frac{n_+F_{s+}-n_-F_{s-}-F_{ext}}{n_+F_{s+}/v_{B+}+n_-F_{s-}/v_{F-}+\gamma}
\end{aligned}
\end{equation}
and the unbinding rates of plus and minus motors are
\begin{equation}\label{eq111}
\begin{aligned}
\epsilon_{+}(x,y)=&\epsilon_{0+}\exp\left(\frac{y
+\bar{a}(F_{ext}+\gamma
v_c)/N_-}{(\bar{a}cx+\bar{b}y)F_{d+}}\right)\cr
\epsilon_{-}(x,y)=&\epsilon_{0-}\exp\left(\frac{cx
-\bar{b}(F_{ext}+\gamma v_c)/N_-}{(\bar{a}cx+\bar{b}y)F_{d-}}\right)
\end{aligned}
\end{equation}

The splitting boundary of these two cases is also
$n_+F_{s+}=n_-F_{s-}+F_{ext}$, i.e.
$y=xcF_{s+}/F_{s-}-F_{ext}/N_-F_{s-}$.

\section{The dynamics of motor numbers $n_+$ and $n_-$}
For the sake of convenience, let
\begin{equation}\label{eq32}
\left\{\begin{aligned} &r_+\equiv r_+(n_+, n_-):=(N_+-n_+)\pi_+\cr &
s_+\equiv s_+(n_+, n_-):=n_+\epsilon_+(n_+, n_-)\cr &r_-\equiv
r_-(n_+, n_-):=(N_--n_-)\pi_-\cr & s_-\equiv s_-(n_+,
n_-):=n_-\epsilon_-(n_+, n_-)
\end{aligned}\right.
\end{equation}
and $\lambda=r_++r_-+s_++s_-$. During time interval $(t, t+\triangle
t)$, the increase of plus motor number is
\begin{equation}\label{eq33}
\begin{aligned}
n_+(t+\triangle
t)-n_+(t)=\left(\frac{r_+}{\lambda}-\frac{s_+}{\lambda}\right)\int_0^{\triangle
t}\lambda e^{-\lambda}dt
=\frac{r_+-s_+}{\lambda}\left(1-e^{-\lambda\triangle t}\right)
\end{aligned}
\end{equation}
In the limit $\triangle t\to 0$, (\ref{eq33}) leads to
\begin{equation}\label{eq34}
\begin{aligned}
\frac{dn_+}{dt}=r_+-s_+=(N_+-n_+)\pi_+-n_+\epsilon_+(n_+, n_-)
\end{aligned}
\end{equation}
Similarly, the dynamics of minus motor number is
\begin{equation}\label{eq35}
\begin{aligned}
\frac{dn_-}{dt}=r_+-s_+=(N_--n_-)\pi_--n_-\epsilon_-(n_+, n_-)
\end{aligned}
\end{equation}
So $x=n_+/N_+,\ y=n_-/N_-$ satisfy
\begin{equation}\label{eq36}
\left\{\begin{aligned}
&\frac{dx}{dt}=\pi_+-x[\pi_++\epsilon_+(x,y)]:=f(x,y)\cr
&\frac{dy}{dt}=\pi_--y[\pi_-+\epsilon_-(x,y)]:=g(x,y)
\end{aligned}\right.
\end{equation}

As we all know, the steady state solutions $(x^*, y^*)$ of the
system (\ref{eq36}), which satisfy $f(x^*,y^*)=0$ and
$g(x^*,y^*)=0$, are stable if and only if the real parts of the two
eigenvalues of the following matrix
\begin{equation}\label{eq362}
\left[\begin{array}{cc} \frac{\partial f}{\partial x}(x^*,
y^*)&\frac{\partial f}{\partial y}(x^*, y^*)\cr \frac{\partial
g}{\partial x}(x^*, y^*)&\frac{\partial g}{\partial y}(x^*, y^*)
\end{array}\right]
\end{equation}
are nonpositive. It is to say that
\begin{equation}\label{eq363}
\begin{aligned}
&\frac{\partial f}{\partial x}(x^*, y^*)+\frac{\partial g}{\partial
y}(x^*, y^*)\le 0\cr &\frac{\partial f}{\partial x}(x^*, y^*)
\frac{\partial g}{\partial y}(x^*, y^*)-\frac{\partial f}{\partial
y}(x^*, y^*) \frac{\partial g}{\partial x}(x^*, y^*)\ge 0
\end{aligned}
\end{equation}

\begin{figure}
  \includegraphics[width=220pt]{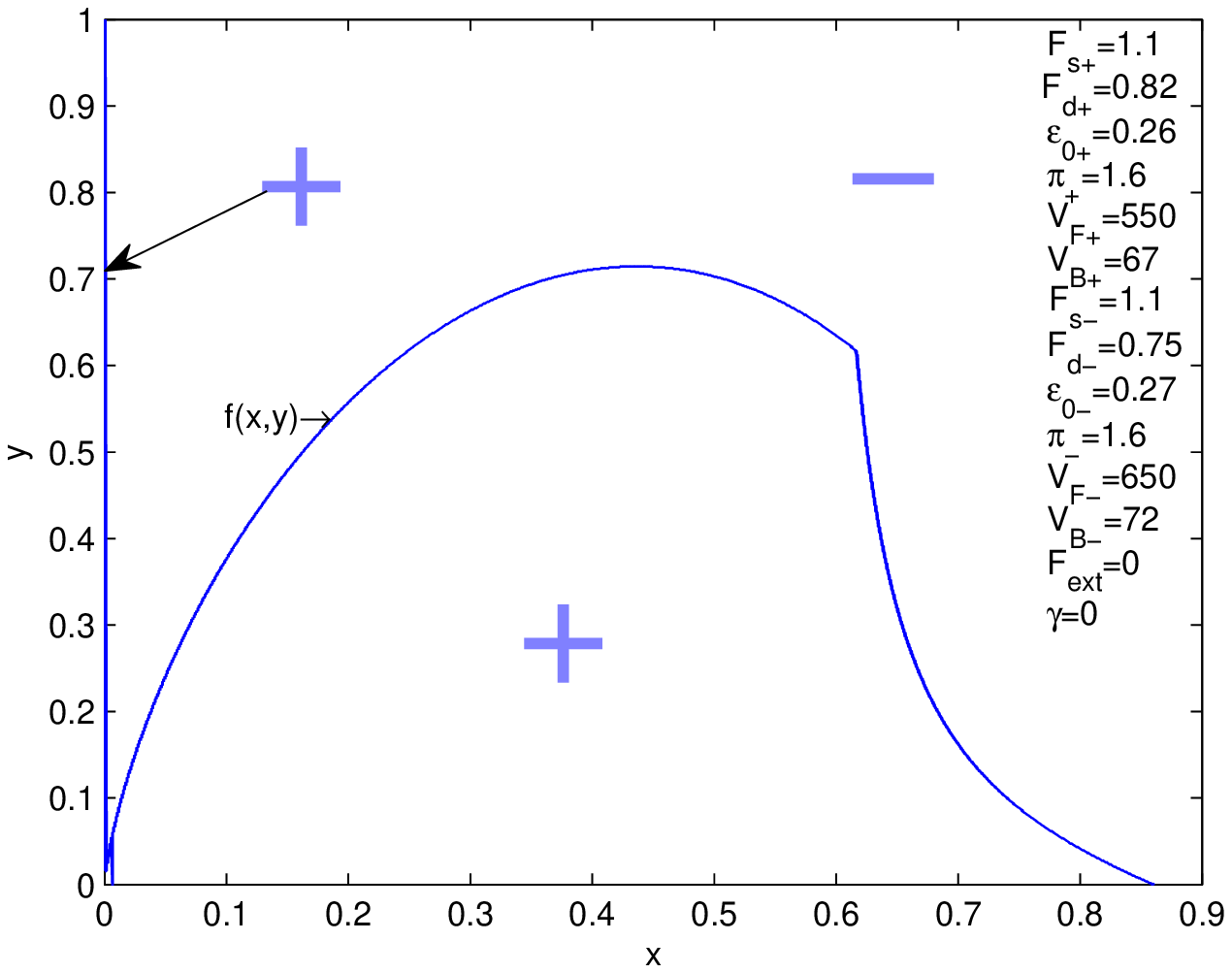}\includegraphics[width=220pt]{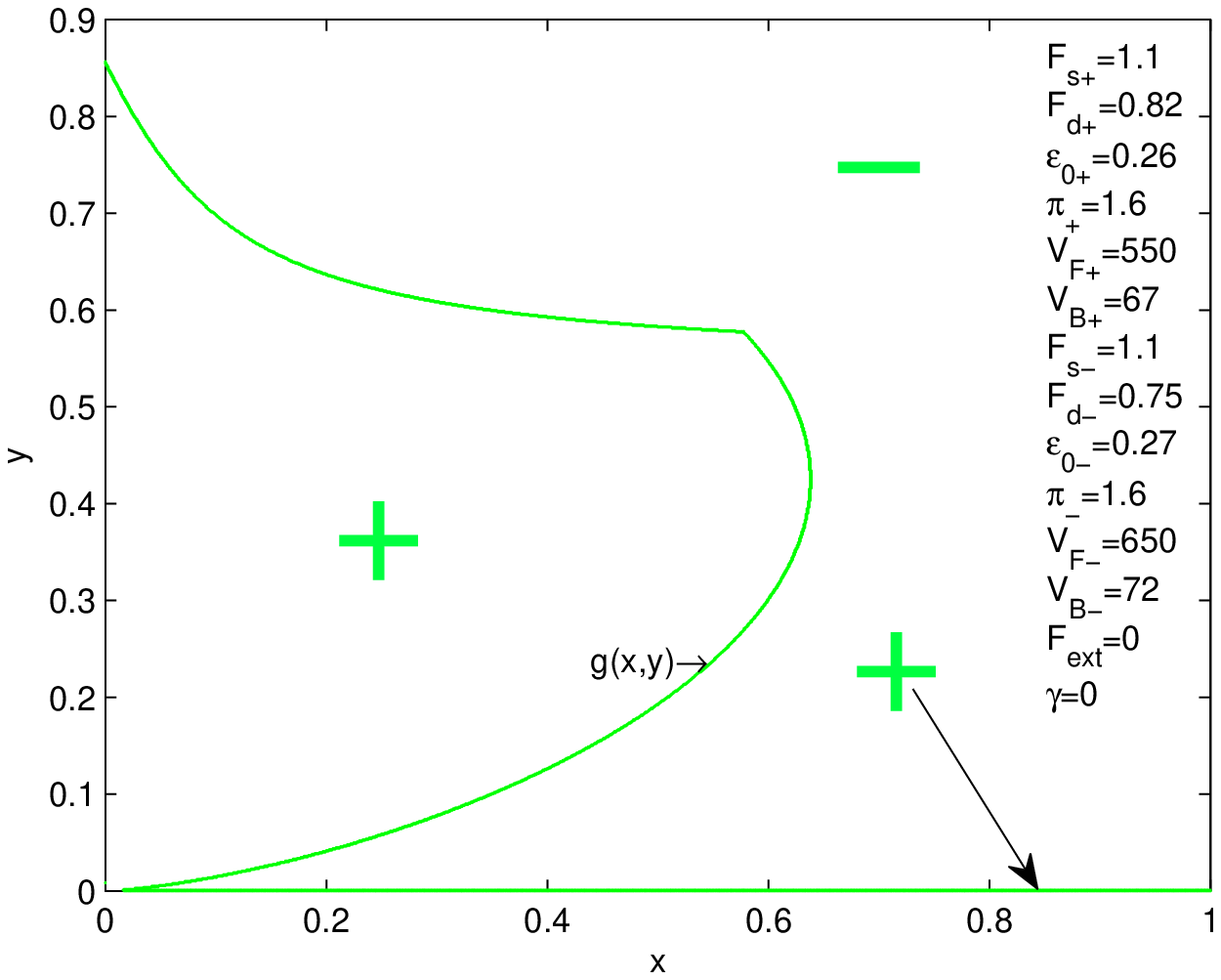}\\
  \caption{The figures of functions $f(x,y)=0,\ g(x,y)=0$. The \lq\lq+" (\lq\lq-") means the function $f$ (or $g$) is positive (negative) in the corresponding
  subdomains.
  }\label{Fig2}
\end{figure}

\begin{figure}
  \includegraphics[width=220pt]{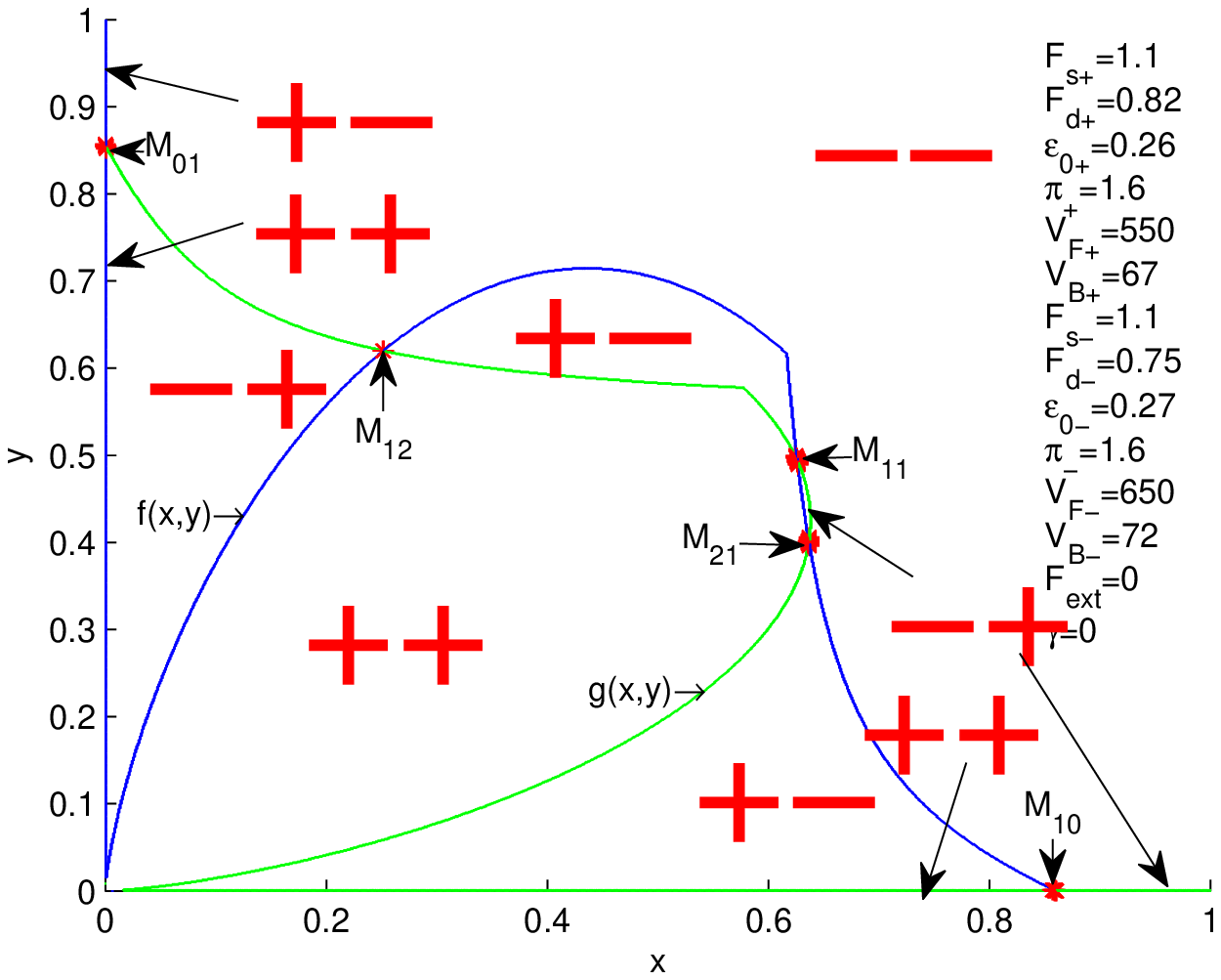}\includegraphics[width=220pt]{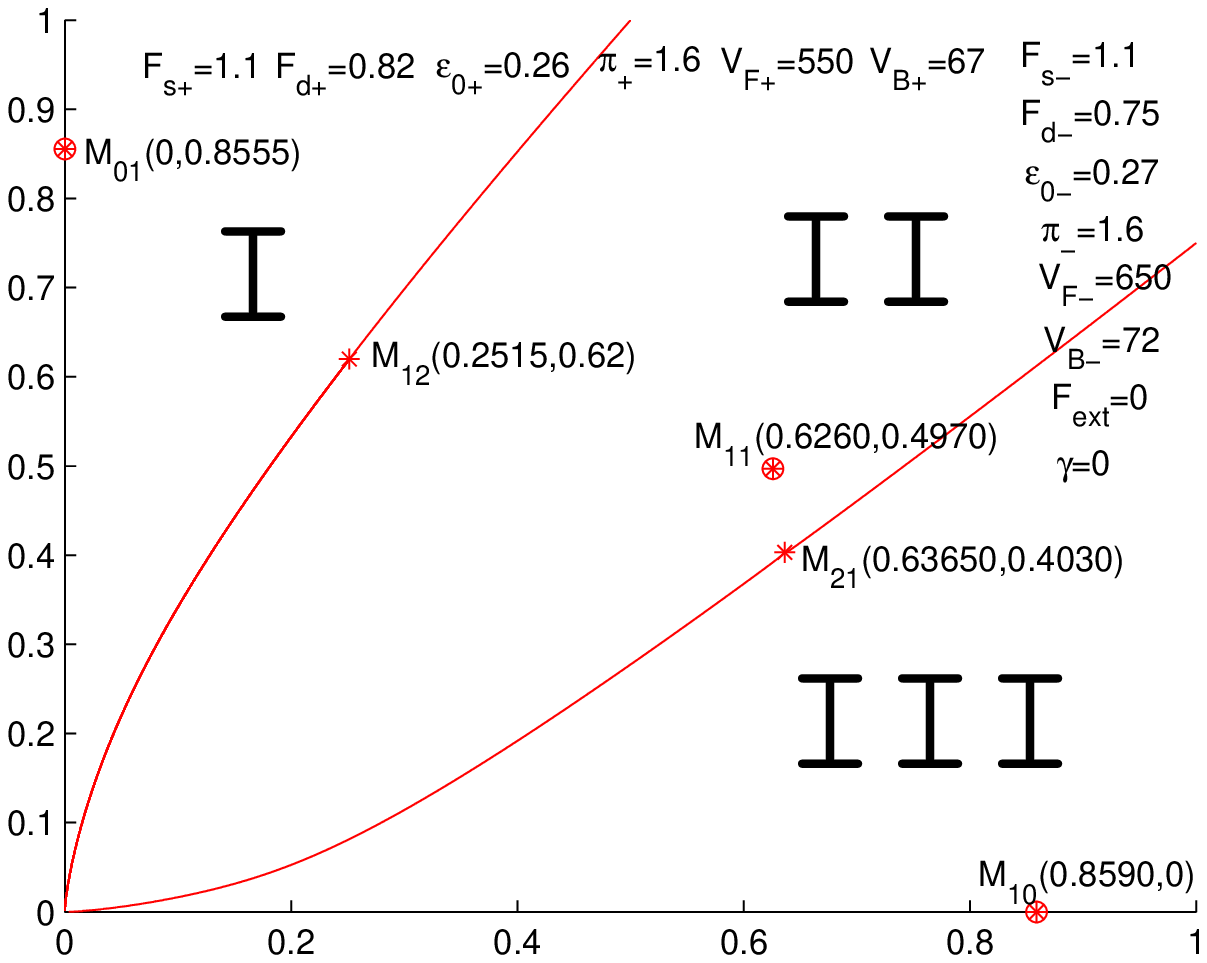}\\
  \caption{The steady states of system (\ref{eq36}). Where the unstable steady
  states are denoted by \lq\lq$*$", the stable steady states are denoted by
  \lq\lq$\circ\hskip-6pt *$". If the initial state $P_0(x_0, y_0)$ lies in the subdomain I (II or III), then the final state is the stable steady state $M_{01}$ ($M_{11}$
  or $M_{10}$).)}\label{Fig3}
\end{figure}

To better understanding, the figures of functions $f(x,y)=0,\
g(x,y)=0$ are plotted in figure \ref{Fig2}.
In view of conditions (\ref{eq363}), to initial values
$x_0=n_+/N_+$, $y_0=n_-/N_-$, if the point $P_0(x_0, y_0)$ lies in
the subdomain I (II or III), then the final state is stable steady
state $M_{01}$ ($M_{11}$ or $M_{10}$) (see Fig. \ref{Fig3}).
Theoretically, $y_{M_{10}}\ne 0$, $x_{M_{01}}\ne 0$, but they are
small than the accuracy of the numerical calculation used in this
paper, so we simply regard them as $0$.

\begin{figure}
  \includegraphics[width=220pt]{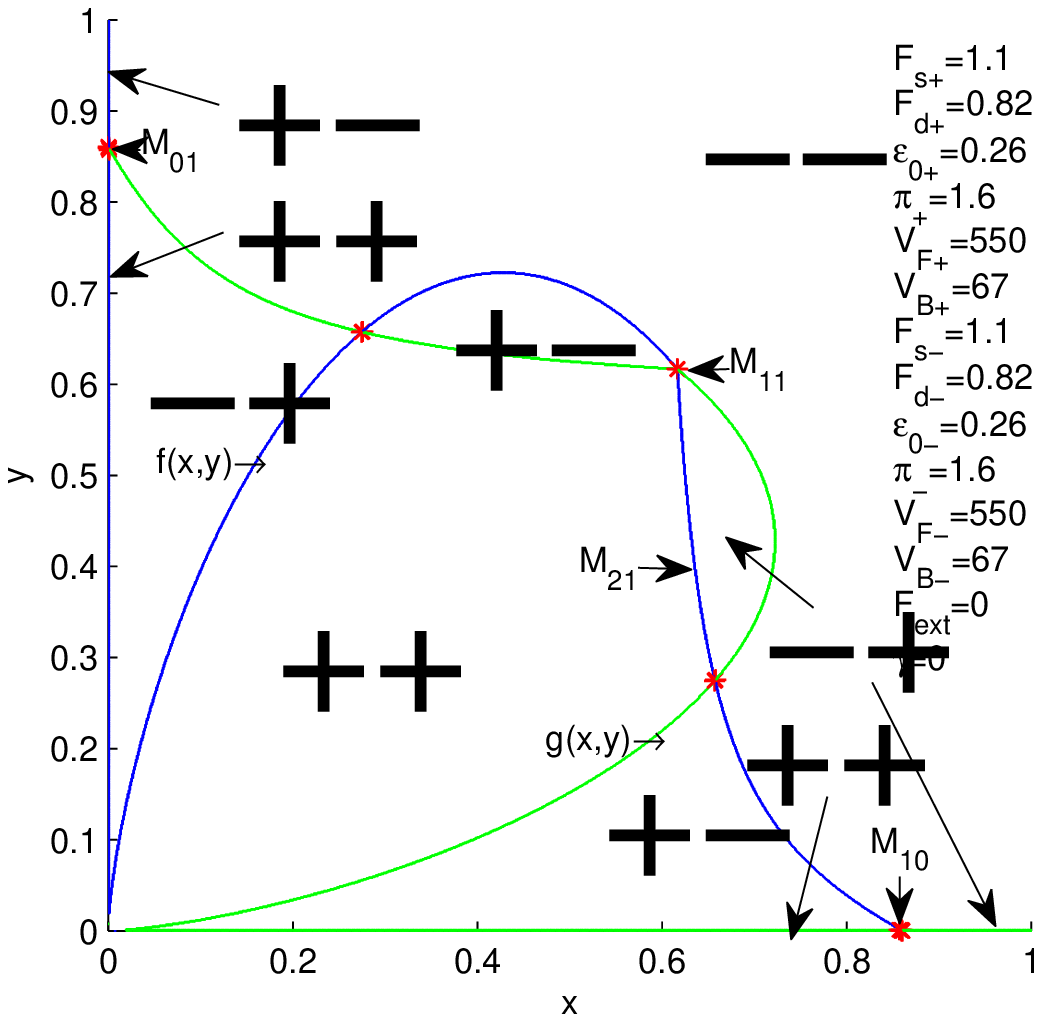}\includegraphics[width=220pt]{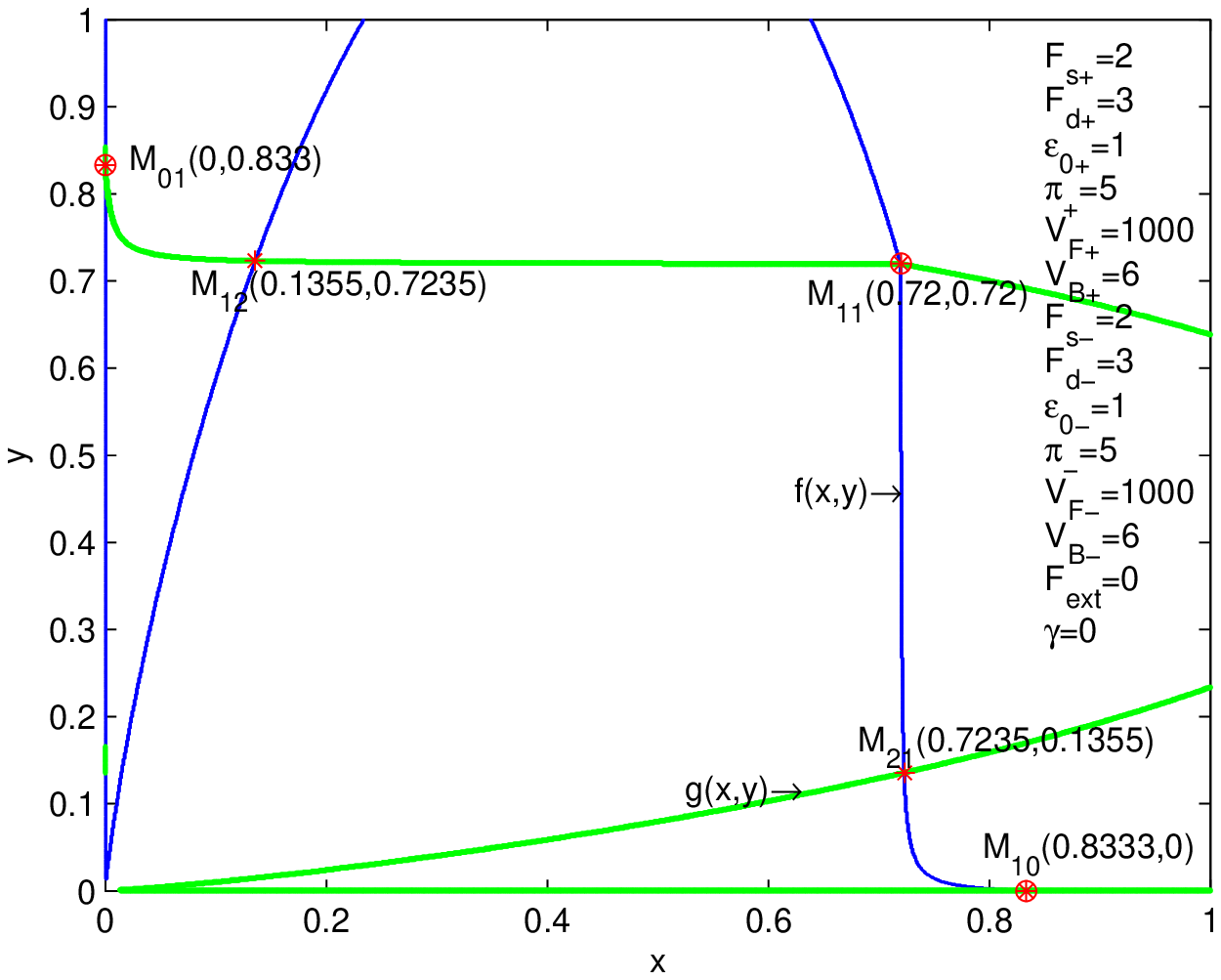}\\
  \includegraphics[width=220pt]{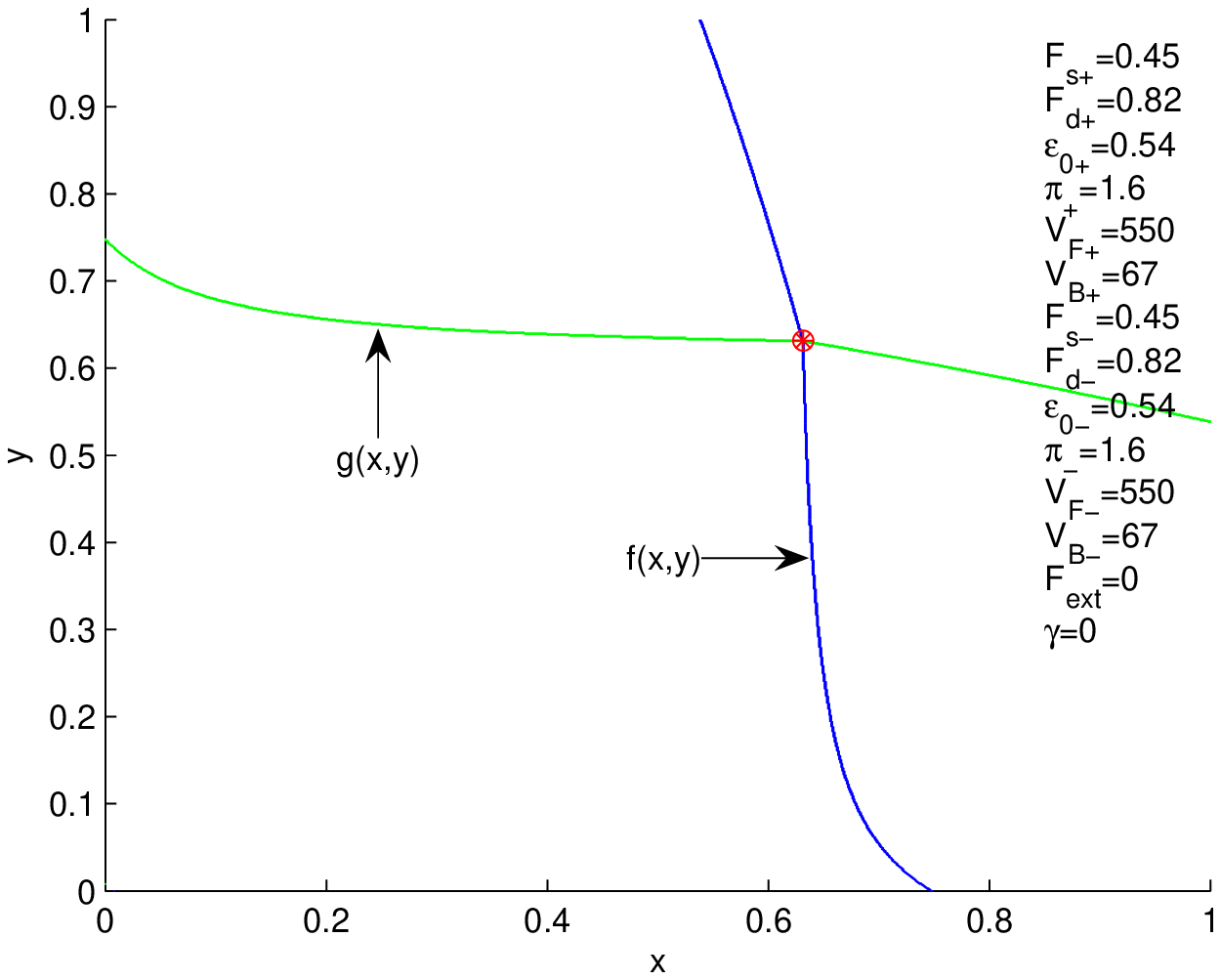}\includegraphics[width=220pt]{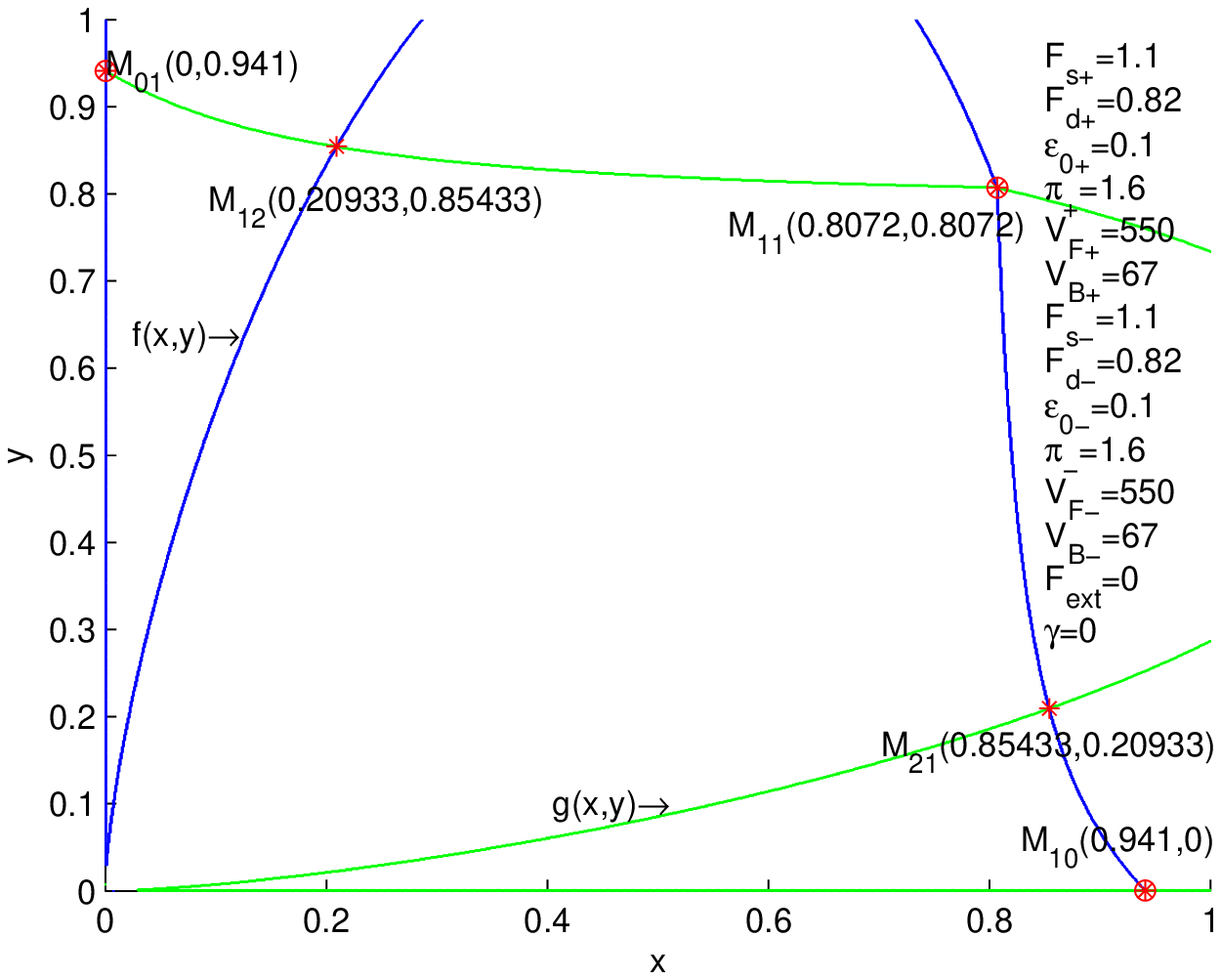}\\
  \caption{Figures of $f(x,y)=0, g(x,y)=0$ for symmetric tug-of-war model, in which plus and minus motors have the same parameters,
   The unstable steady states are denoted by \lq\lq$*$", the stable steady states are denoted by
  \lq\lq$\circ\hskip-6pt *$".}\label{Fig4}
\end{figure}

\begin{figure}
  \includegraphics[width=150pt]{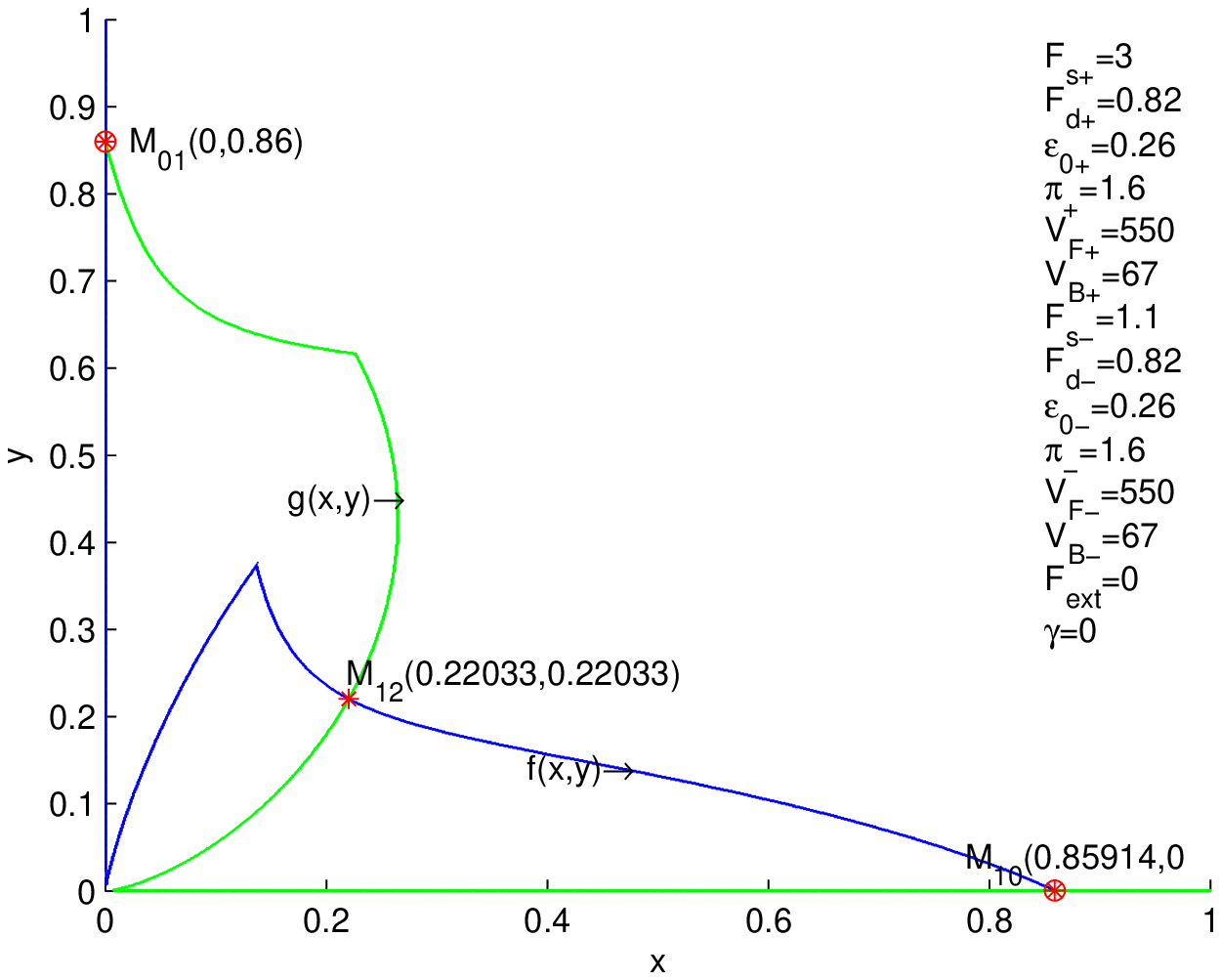}\includegraphics[width=150pt]{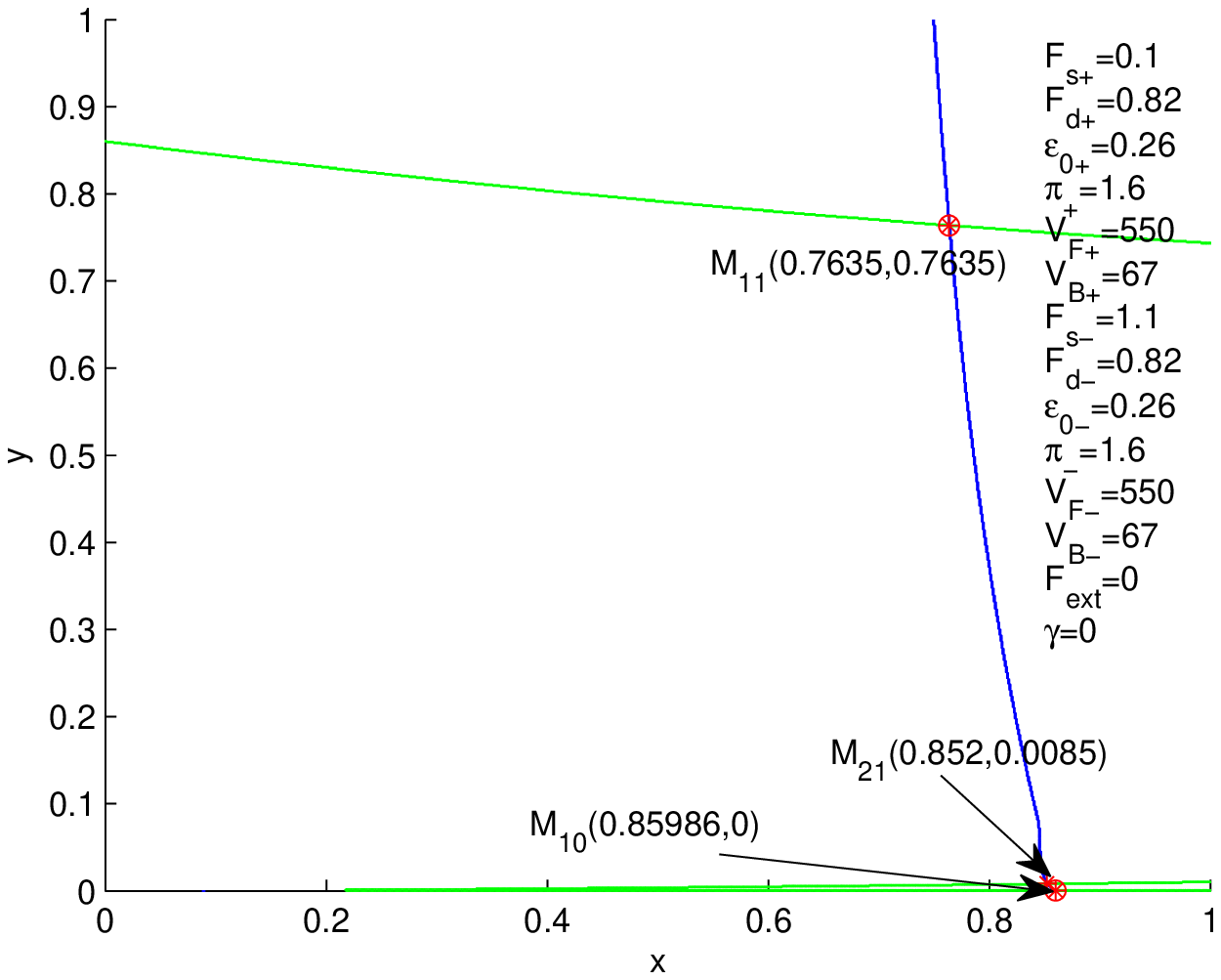}\includegraphics[width=150pt]{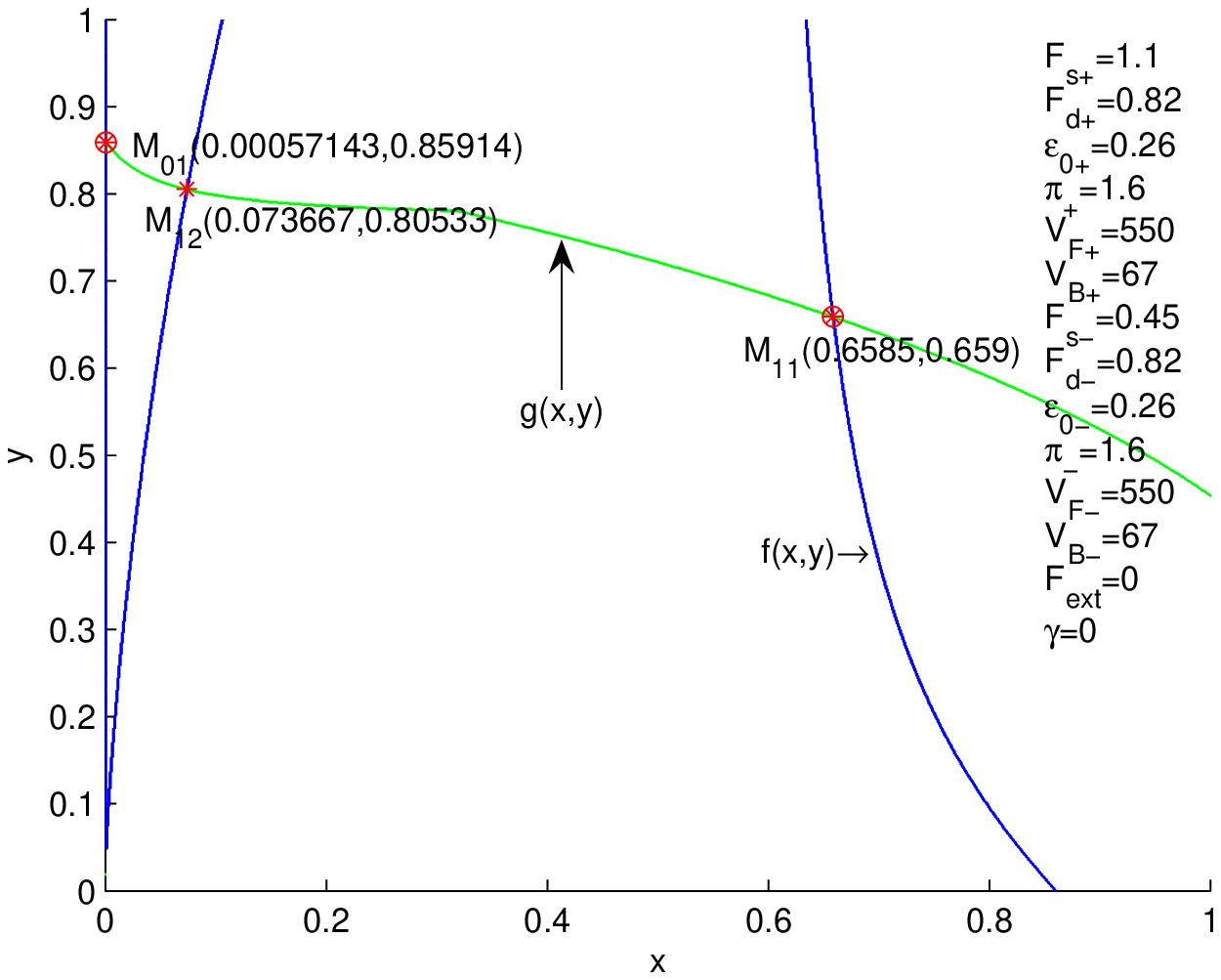}\\
  \caption{Asymmetric tug-of-war model: In this case, the system (\ref{eq36})
  might have one, two or three stable steady states.}\label{Fig5}
\end{figure}

\begin{figure}
  \includegraphics[width=220pt]{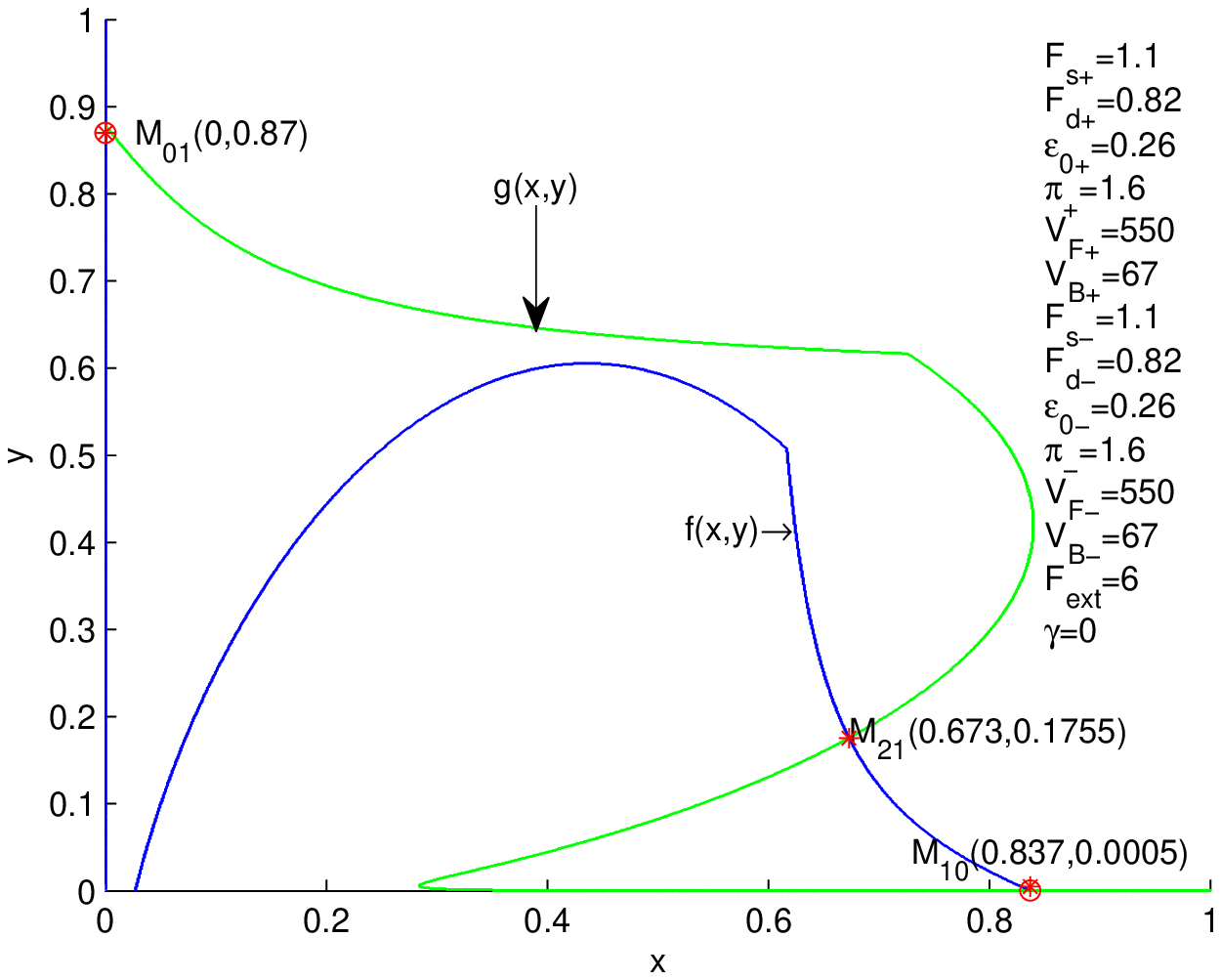}\includegraphics[width=220pt]{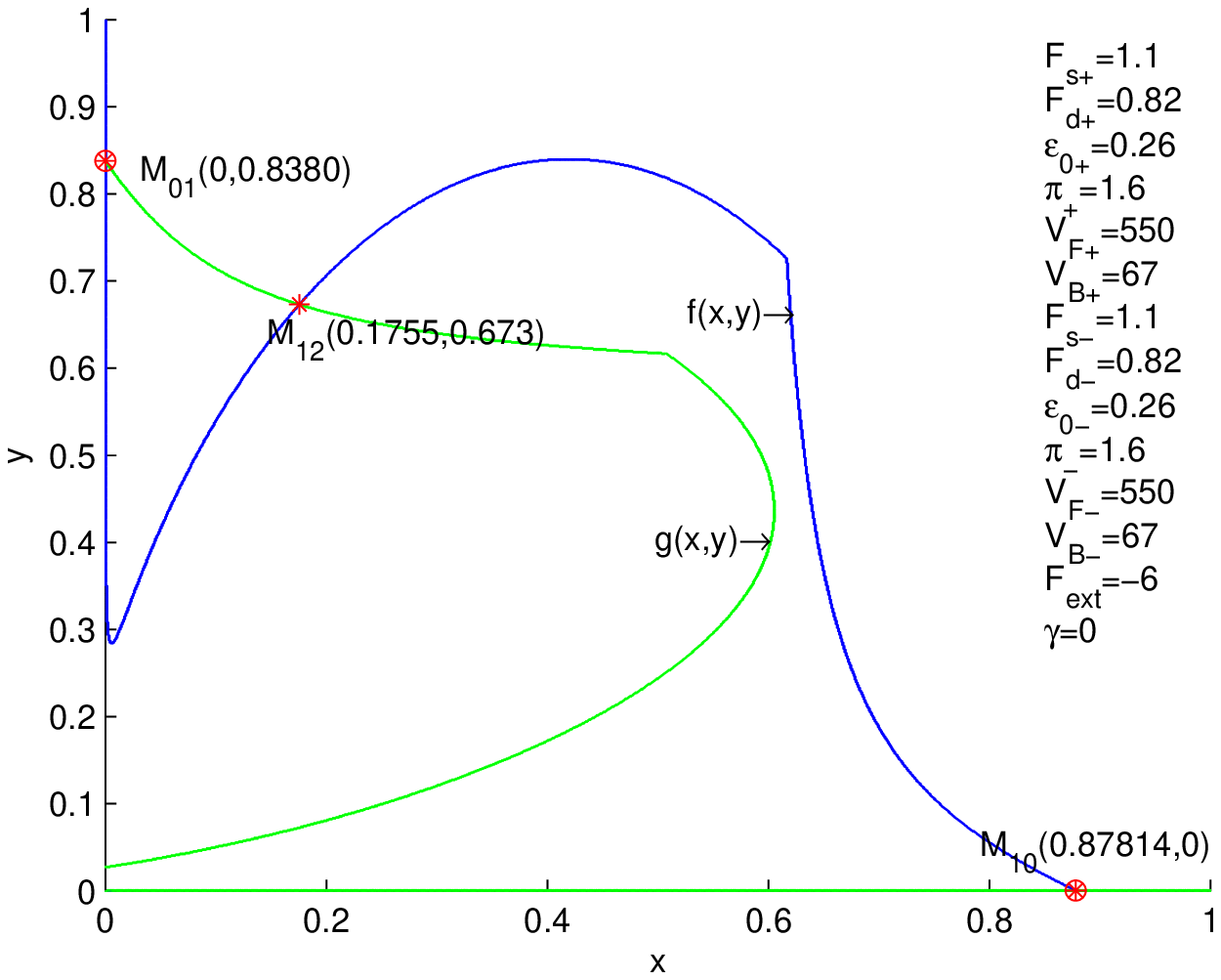}\\
  \includegraphics[width=220pt]{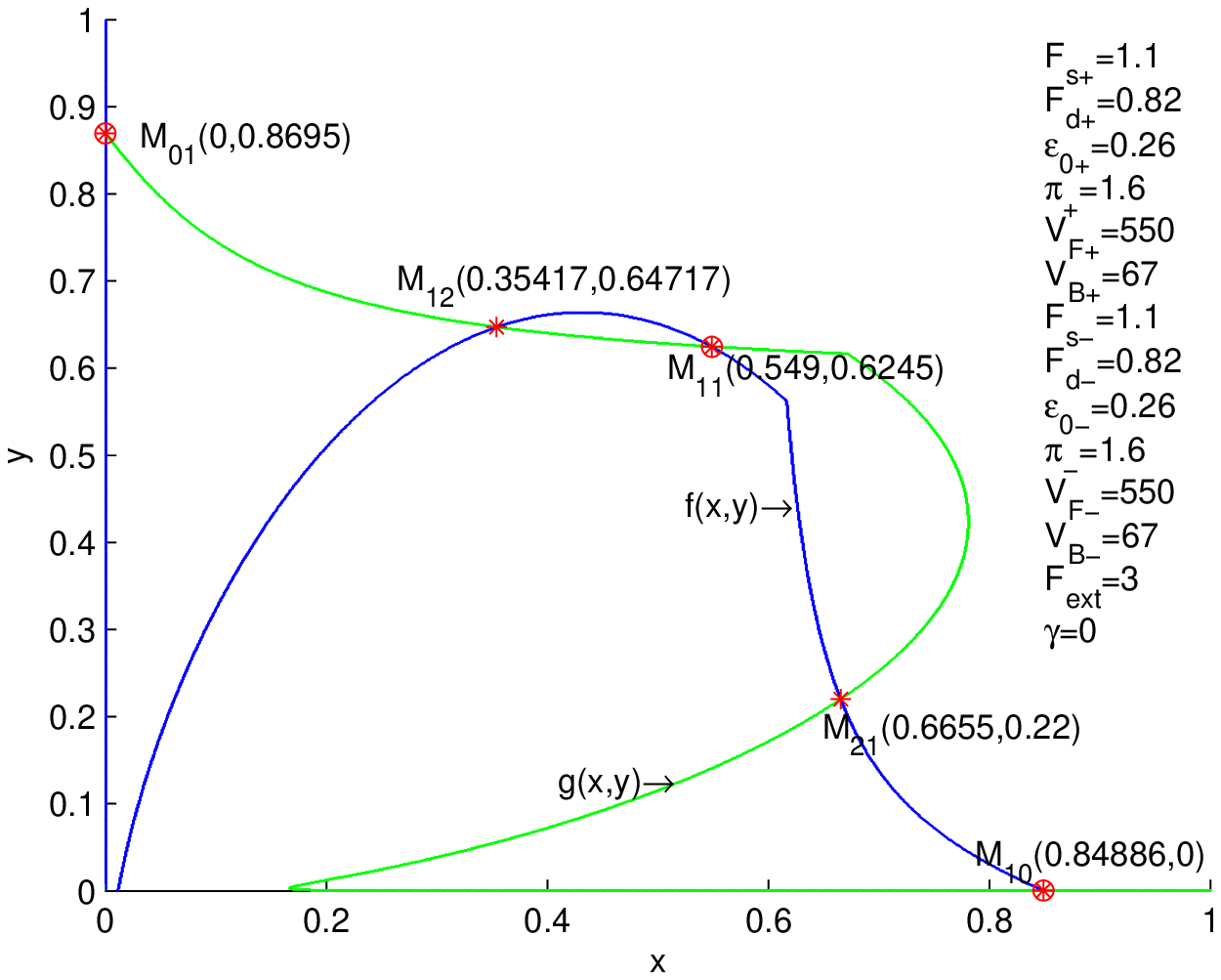}\includegraphics[width=220pt]{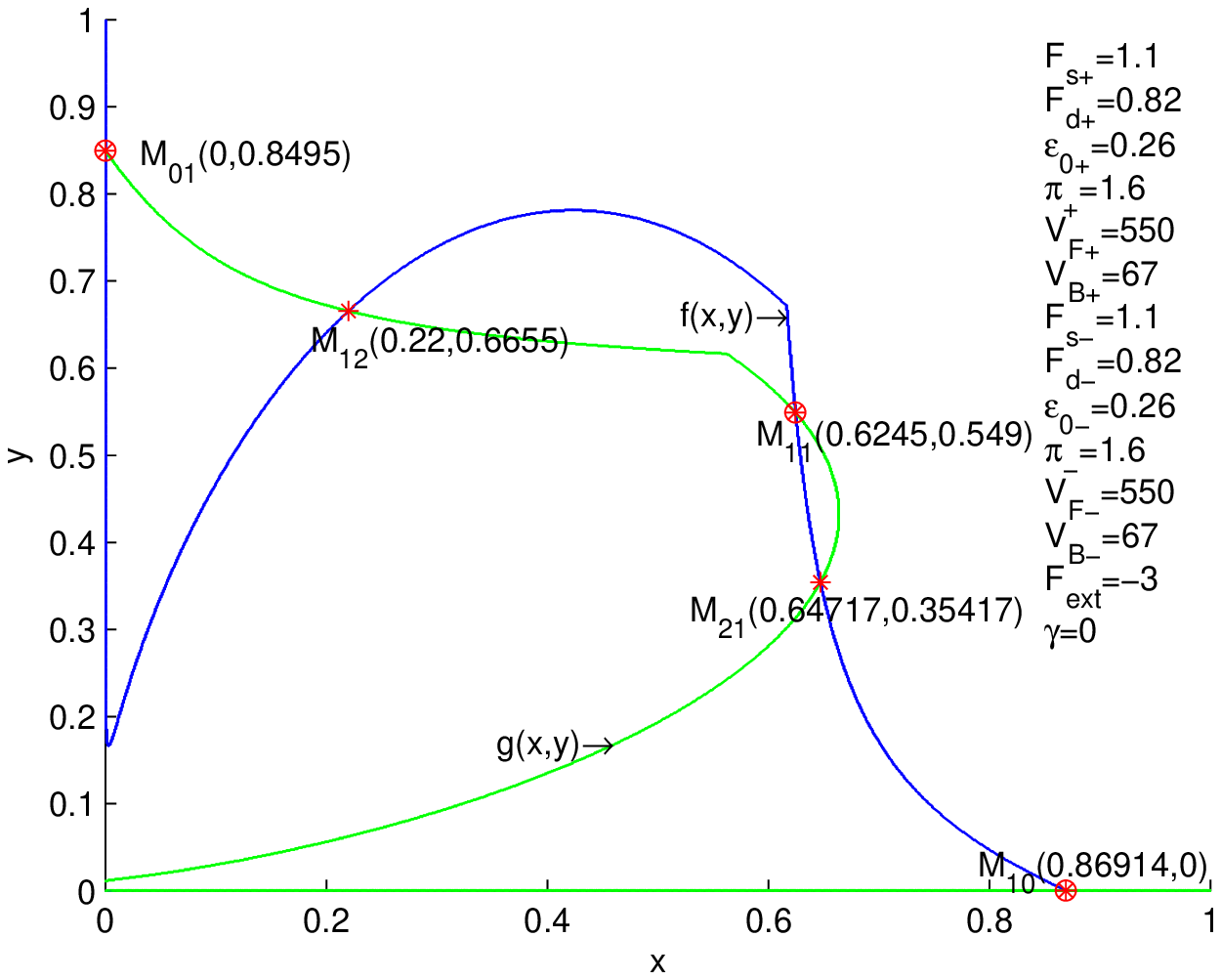}\\
  \caption{Tug-of-war model with external force $F_{ext}$: In this case, the system (\ref{eq36})
  might have two or three stable steady states.}\label{Fig6}
\end{figure}

To further understand the properties of the stable steady state
points, the figures of $f(x,y)=0$ and $g(x,y)=0$ with different
values of parameters $F_{s+}, F_{s-}$, $F_{d+}, F_{d-}$, $v_{B+}$,
$v_{B-}$, $v_{F+}$, $v_{F-}$, $\pi_{+}, \pi_{-}$, $\epsilon_{0+},
\epsilon_{0-}$ and $c=N_+/N_-$ are plotted in Fig. \ref{Fig4} and
\ref{Fig5}, \ref{Fig6}.
From the figures, one can find that system (\ref{eq36}) might have
one, two or three stable steady states, which depends on the values
of the parameters. Given the initial value $(x_0, y_0)$, the final
steady state can be determined using the similar method as in Fig.
\ref{Fig3} ({\bf Right}). One can be easily know that, almost all of
the parameters used in the tug-of-war model have one or two critical
points, the final stable steady state would change if one of the
parameters jumps from one side of its critical points to another
side.

Obviously, for $N_+=0$ or $N_-=0$ (i.e. $c=0$ or $c=\infty$), the
tug-of-war model is reduced to the usual model for cooperate
transport by a single motor species (minus or plus), and the only
stable steady state is $\pi_+/(\pi_++\epsilon_{0+})$ for plus motor
species or $\pi_-/(\pi_-+\epsilon_{0-})$ for minus motor species.
The average velocity of the cargo at steady state is
$v_c=v_c(x^*,0)=v_{F+}$ if $c=\infty$, and $v_c=v_c(0,y^*)=-v_{B-}$
if $c=0$, which are the velocities of a single motor.

\section{Comparison with Monte Carlo simulations}
\begin{figure}
  \includegraphics[width=150pt]{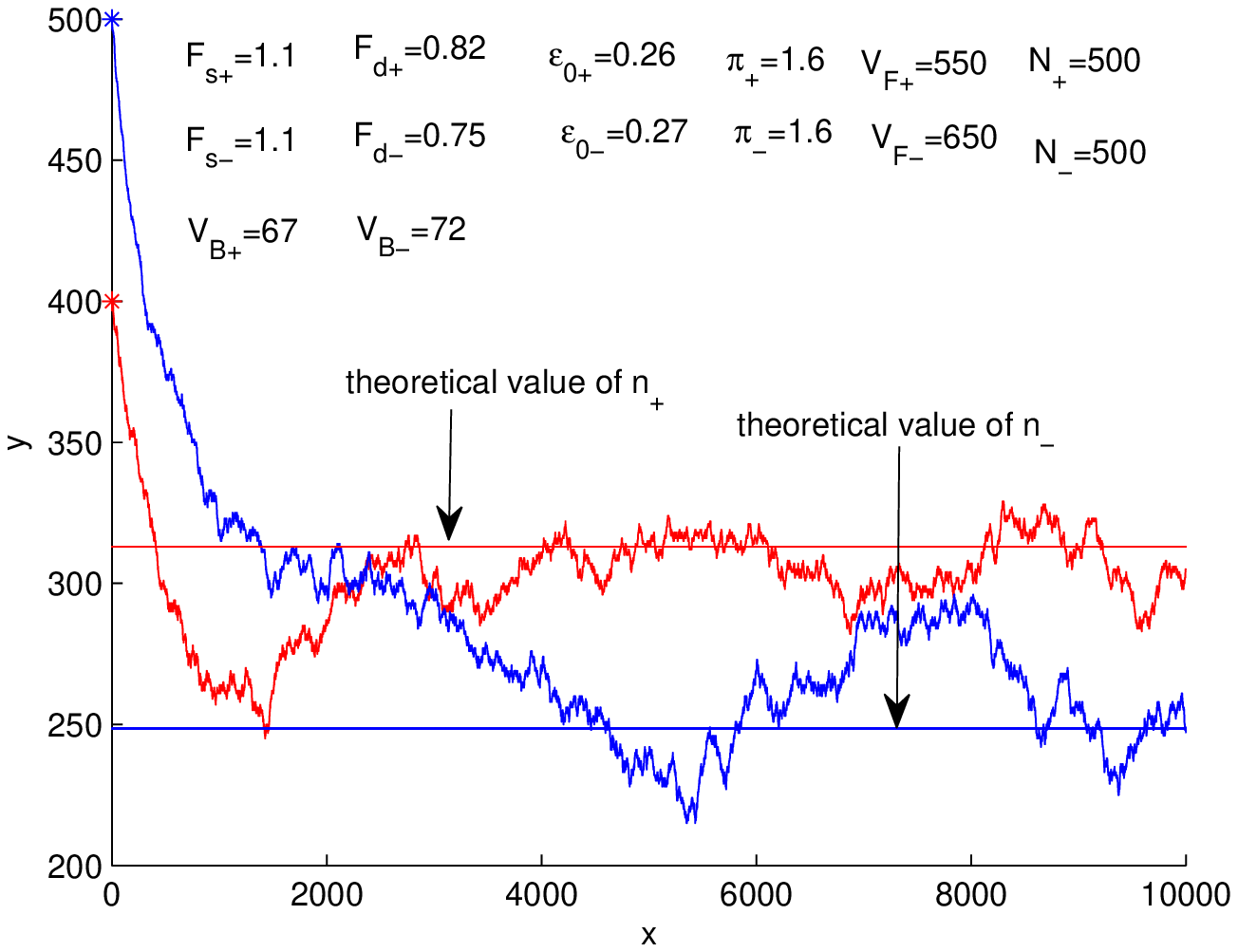}\includegraphics[width=150pt]{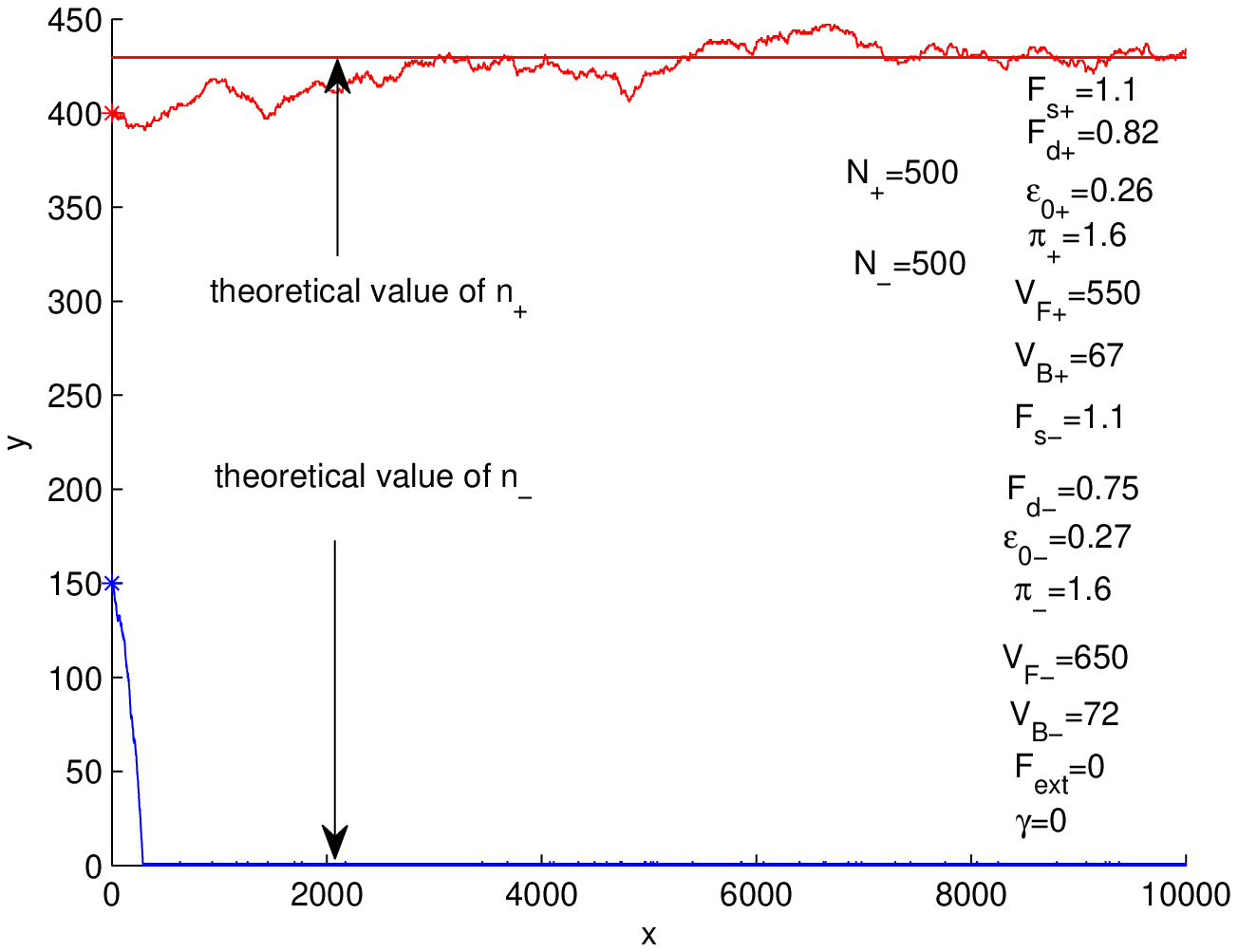}\includegraphics[width=150pt]{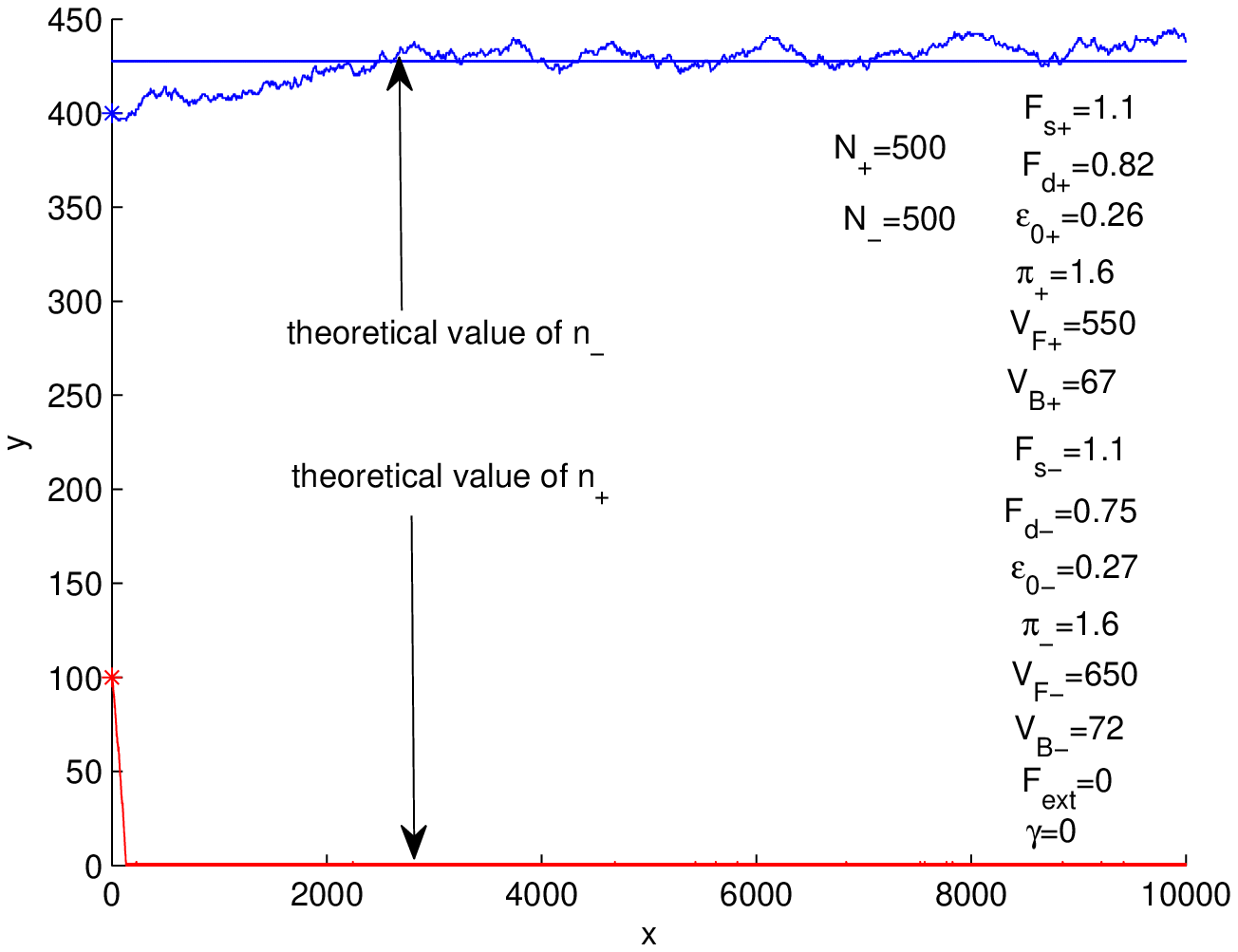}\\
  \caption{For large motor numbers $N_+, N_-$ cases, the steady
  states is determined by the theoretical steady state
  $n_{+}^s\approx N_+x^s, n_{-}^s\approx N_-y^s$. {\bf Left: }$(n_+(0)/N_+, n_-(0)/N_-)$ lie in subdomain ({\bf
  II});   {\bf Middle: }$(n_+(0)/N_+, n_-(0)/N_-)$ lie in subdomain ({\bf III});
   {\bf Right: } $(n_+(0)/N_+, n_-(0)/N_-)$ lie in subdomain ({\bf I}).}\label{Fig7}
\end{figure}

Due to the above discussion, in large motor numbers limit $N_+,
N_-\to\infty$, the movement of the cargo might have one, two or
three stable steady states. The final steady state is determined by
the initial motor numbers $n_+(0)=N_+x_0$ and $n_-(0)=N_-y_0$ (see
Fig. \ref{Fig7}).
For example, in case of Fig. \ref{Fig3} (right), if $(n_+(0)/N_+,
n_-(0)/N_-)$ lies in subdomains {\bf (II) }, the final steady state
would be $n^s_+\approx N_+x_{M_{11}}, n^s_-\approx N_-y_{M_{11}}$.

\begin{figure}
  \includegraphics[width=220pt]{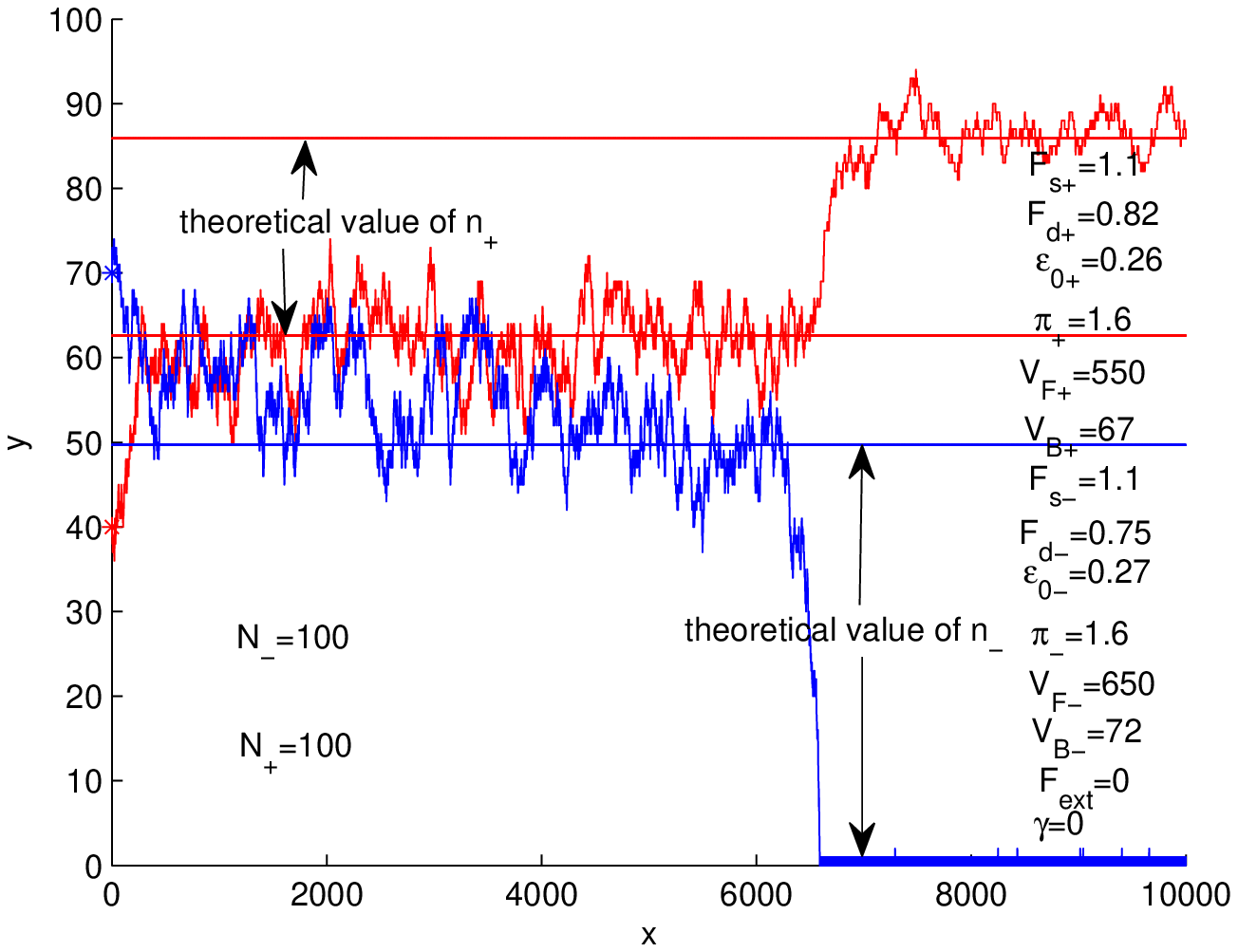}\includegraphics[width=220pt]{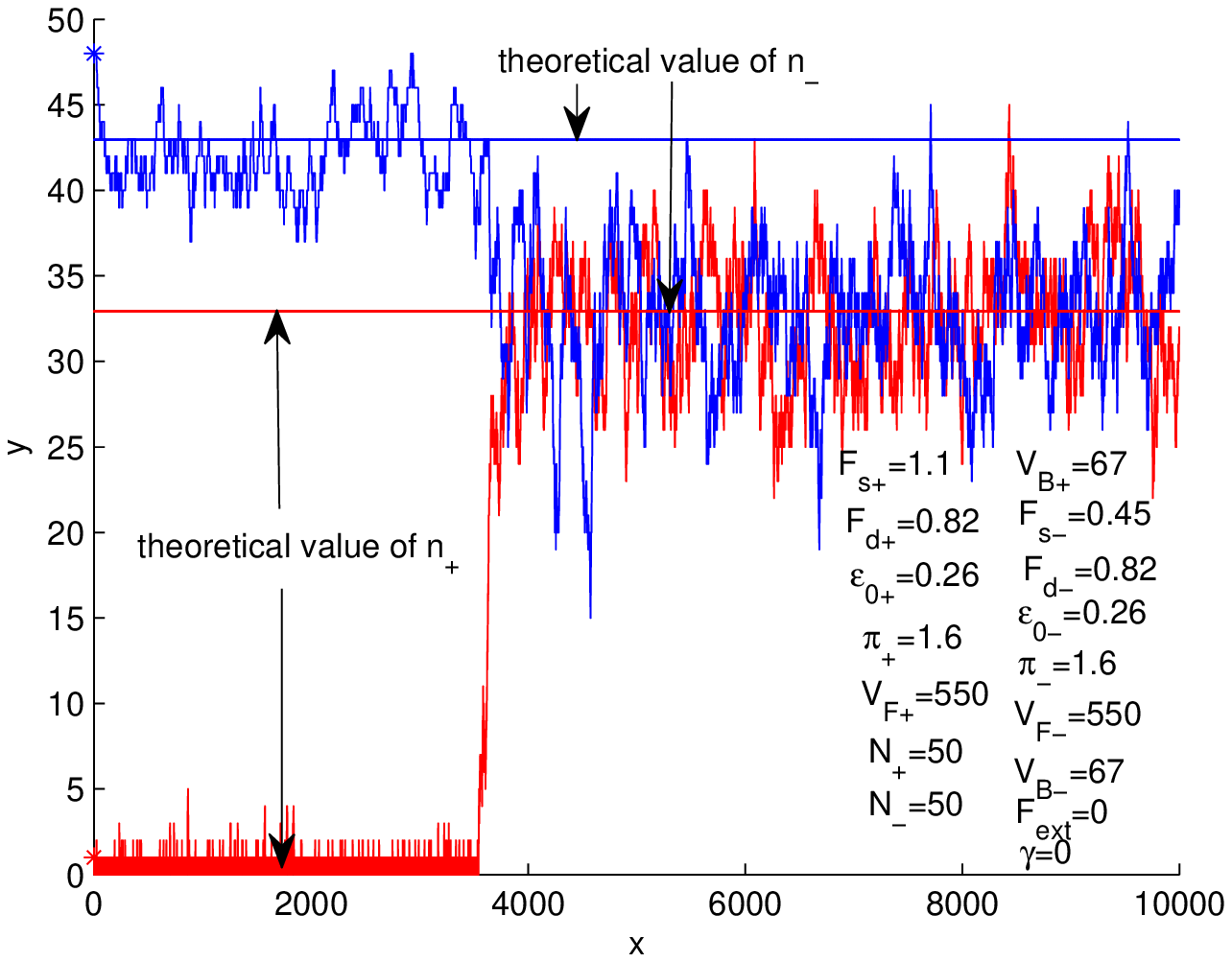}\\
  \caption{For small motor numbers $N_+, N_-$, the final motor numbers $n_+, n_-$ can change from one stable steady state
  to another. {\bf Left: }The final motor numbers $n_+, n_-$ change from $N_+x_{M_{11}}$, $N_-y_{M_{11}}$ to
  $N_+x_{M_{10}}$, $N_-y_{M_{10}}$;  {\bf Right: }The final motor numbers $n_+, n_-$ change from $N_+x_{M_{01}}$, $N_-y_{M_{01}}$ to
  $N_+x_{M_{11}}$, $N_-y_{M_{11}}$.}\label{Fig8}
\end{figure}

However, if the numbers $N_+, N_-$ of molecular motors, which
attached to the cargo, is finite or even small, the steady states
numbers $n^s_+$ and $n^s_-$ might be different with the theoretical
values $N_+x^*$ and $N_-y^*$. Theoretically, if $M_{i}(x_{i}, y_{i})
(i=1, 2 \textrm{ or } 3)$ are the stable steady points of the system
(\ref{eq36}), which can be regarded as the large motor numbers limit
of (\ref{eq34}, \ref{eq35}), then steady state numbers $n^s_+$ and
$n^s_-$ would lie in the neighborhoods of the theoretical values
$N_+x_{i}$ and $N_-y_{i}$. But, in small $N_+, N_-$ cases, the
steady state motor numbers $n^s_+$ and $n^s_-$ can jump easily from
the neighborhood of one of the theoretical stable steady state point
$(N_+x_{i}, N_-y_{i})$ to the neighborhood of another theoretical
stable steady state point $(N_+x_{j}, N_-y_{j})$ (see Fig.
\ref{Fig8}).
For finite motor numbers $N_+, N_-$, the stepsize of the system
(\ref{eq36}) are $\triangle x=1/N_+$, $\triangle y=1/N_-$. So the
smaller of motor numbers $N_+, N_-$, the easier for motor numbers
$n_+, n_-$ to jump from one of the steady subdomains {\bf I, II
\textrm{or} III} to another. Intuitively, the probability that
$(n_+/N_+, n_-/N_-)$ lies in the neighborhood of the stable steady
state point $M_{i}$ is proportional to the area of $M_{i}$'s steady
subdomain. Mathematically, the probability of motor numbers $n_+,
n_-$ change from $n_+^{(1)}, n_-^{(1)}$ to $n_+^{(2)}, n_-^{(2)}$
along trajectory $S$ is
\begin{equation}\label{eq29}
\begin{aligned}
p^{12}_S=&\left[\prod_{(S_i, S_{i+1})\in
S_R}\frac{\pi_{i+}}{\pi_{i+}+\epsilon_{i+}+\pi_{i-}+\epsilon_{i-}}\right]\left[\prod_{(S_j,
S_{j+1})\in
S_L}\frac{\epsilon_{j+}}{\pi_{j+}+\epsilon_{j+}+\pi_{j-}+\epsilon_{j-}}\right]\cr
&\left[\prod_{(S_k, S_{k+1})\in
S_U}\frac{\pi_{k-}}{\pi_{k+}+\epsilon_{k+}+\pi_{k-}+\epsilon_{k-}}\right]\left[\prod_{(S_l,
S_{l+1})\in
S_D}\frac{\epsilon_{l-}}{\pi_{l+}+\epsilon_{l+}+\pi_{l-}+\epsilon_{l-}}\right]
\end{aligned}
\end{equation}
where $S_L\cup S_R\cup S_U\cup S_D=S$, $(P_1, P_2)\in S_R$ if and
only if $n_+(P_2)=n_+(P_1)+1, n_-(P_2)=n_-(P_1)$, $(P_1, P_2)\in
S_L$ if and only if $n_+(P_2)=n_+(P_1)-1, n_-(P_2)=n_-(P_1)$, $(P_1,
P_2)\in S_U$ if and only if $n_+(P_2)=n_+(P_1),
n_-(P_2)=n_-(P_1)+1$, $(P_1, P_2)\in S_D$ if and only if
$n_+(P_2)=n_+(P_1), n_-(P_2)=n_-(P_1)-1$. So, theoretically, we can
obtain the probability that motor numbers $n_+, n_-$ change from the
neighborhood of one stable steady states to the neighborhood of
another stable steady states. From these transition probabilities,
we can know more details about the steady state movement of the
cargo in this small $N_+, N_-$ cases.

\section{Conclusion and remarks}
In this paper, the steady state properties of the recent tug-of-war
model, which is provided by Lipowsky {\it et al} to model the
movement of cargo, which is transported by two motor species in the
cell, is discussed. Biophysically, the stable steady states are the
most important states, because the transition time to the stable
steady state, as illustrated in this paper, is very short (see Figs.
\ref{Fig7} and \ref{Fig8}), so almost all of the data are measured
in stable steady states. Through the discussion in this paper, we
can know that the final steady states of the movement of the cargo
is determined by initial numbers of the plus and minus motors which
are bounded to the microtubule. Certainly, the velocity and
direction of the movement are also determined by other several
parameters, such as $N_{\pm},
F_{s\pm}, \pi_{\pm}, \epsilon_{0\pm}, F_{d\pm}, v_{F\pm}, v_{B\pm}, F_{ext}, \gamma$. %the total numbers of
%the forward and backward motors which are bounded to the cargos.
One can also find that, almost each of the parameters has critical
points, which determine the stable steady velocity and direction of
the cargo. It is most probable that, many of the parameters,
including the numbers $N_+$ and $N_-$ of plus and minus motors which
are tightly attached to the cargo, and the initial binding numbers
$n_+(0)$ and $n_-(0)$, can be determined by the biochemical
environment and properties of the cargos, so some of which can be
transported from the plus end to the minus end, and others can be
transported reversely.

\vskip 0.5cm

\noindent{\bf Acknowledgments}

This work was funded by National Natural Science Foundation of China
(Grant No. 10701029). The author thanks professor Hong Qian of
University of Washington for his help to complete this research. The
author also thanks the reviewers for their help to improve the
quality of this paper.


\begin{thebibliography}{10}

\bibitem{David2007}
David~D. Hackney.
\newblock Processive motor movement.
\newblock {\em Science}, 316:58--59, 2007.

\bibitem{Carter2005}
N.~J. Carter and R.~A. Cross.
\newblock Mechanics of the kinesin step.
\newblock {\em Nature}, 435:308--312, 2005.

\bibitem{Taniguchi2005}
Y.~Taniguchi, M.~Nishiyama, Y.~Ishhi, and T.~Yanagida.
\newblock Entropy rectifies the brownian step of kinesin.
\newblock {\em Nature Chemical Biology}, 1:342--347, 2005.

\bibitem{Zhang20081}
Yunxin Zhang.
\newblock Three phase model of the processive motor protein kinesin.
\newblock {\em Biophysical Chemistry}, 136:19--22, 2008.

\bibitem{Vale2003}
R.~D. Vale.
\newblock The molecular motor toolbox for intracellular transport.
\newblock {\em Cell}, 112:467--480, 2003.

\bibitem{Sakakibara1999}
H.~Sakakibara, H.~Kojima, Y.~Sakai, E.~Katayama, and K.~Oiwa.
\newblock Inner-arm dynein c of chlamydomonas flagella is a single-headed
  processive motor.
\newblock {\em Nature}, 400:596--589, 1999.

\bibitem{Hooft2007}
A.~M. Hooft, E.~J. Maki, K.~K. Cox, and J.~E. Baker.
\newblock An accelerated state of myosin-based actin motility.
\newblock {\em Biochemistry}, 46:3513--3520, 2007.

\bibitem{Christof2006}
J.~Christof, M.~Gebhardt, Anabel E.-M. Clemen, Johann Jaud, and
Matthias Rief.
\newblock Myosin-v is a mechanical ratchet.
\newblock {\em Proceedings of the National Academy of Sciences of the United
  States of America}, 103:8680--8685, 2006.

\bibitem{Katsuyuki2007}
Katsuyuki Shiroguchi and Kazuhiko~Kinosita Jr.
\newblock Myosin v walks by lever brownian motion.
\newblock {\em Science}, 316:1208--1212, 2007.

\bibitem{Noji1997}
H.~Noji, R.~Yasuda, M.~Yoshida, and Jr.~K. Kinosita.
\newblock Direct observation of the rotation of f1-atpase.
\newblock {\em Nature}, 386:299--302, 1997.

\bibitem{Howard2001}
J.~Howard.
\newblock {\em Mechanics of Motor Proteins and the Cytoskeleton}.
\newblock Sinauer Associates, Sunderland, MA, 2001.

\bibitem{Wang1998}
Hongyun Wang and George Oster.
\newblock Energy transduction in the f1 motor ofatp synthase.
\newblock {\em Nature}, 396:279--280, 1998.

\bibitem{Masayoshi2002}
Masayoshi Nishiyama, Hideo Higuchi, and Toshio Yanagida.
\newblock Chemomechanical coupling of the forward and backward steps of single
  kinesin molecules.
\newblock {\em Nature Cell Biology}, 4:790--797, 2002.

\bibitem{Wang2005}
Hongyun Wang.
\newblock Chemical and mechanical efficiencies of molecular motors and
  implications for motor mechanisms.
\newblock {\em Journal of Physics: Condensed Matter}, 17:S3997--S4014, 2005.

\bibitem{Zhang2008}
Yunxin Zhang.
\newblock The efficiency of the molecular motors.
\newblock {\em Journal of Statistical Physics}, 134:669--679, 2009.

\bibitem{voboda1994}
K.~Svoboda and S.M. Block.
\newblock Force and velocity measured for single kinesin molecules.
\newblock {\em Cell}, 77:773--784, 1994.

\bibitem{Reimann2002}
Peter Reimann.
\newblock Brownian motors: noisy transport far from equilibrium.
\newblock {\em Physics Reports}, 361:57, 2002.

\bibitem{Wang2008}
Hongyun Wang.
\newblock Several issues in modeling molecular motors.
\newblock {\em Journal of Computational and Theoretical Nanoscience}, 5:1--35,
  2008.

\bibitem{Qian2000}
Hong Qian.
\newblock The mathematical theory of molecular motor movement and
  chemomechanical energy transduction.
\newblock {\em Journal of Mathematical Chemistry}, 27:219--234, 2000.

\bibitem{Reimann2001}
P.~Reimann, C.~Van den Broeck, P.~Hanggi H.~Linke, J.M. Rubi, and
  A.~P\'{e}rez-Madrid.
\newblock Giant acceleration of free diffusion by use of tilted periodic
  potentials.
\newblock {\em Physical Review Letters}, 87:010602, 2001.

\bibitem{Astumian1997}
R.D. Astumian.
\newblock Thermodynamics and kinetics of a brownian motor.
\newblock {\em Science}, 276:917--922, 1997.

\bibitem{Astumian2005}
R~Dean Astumian.
\newblock Biasing the random walk of a molecular motor.
\newblock {\em Journal of Physics: Condensed Matter}, 17:S3753--S3766, 2005.

\bibitem{Liepelt2007}
Steffen Liepelt and Reinhard Lipowsky.
\newblock Kinesin¡¯s network of chemomechanical motor cycles.
\newblock {\em Physical Review Letters}, 98(25):258102, 2007.

\bibitem{Fisher2001}
Michael~E. Fisher and Anatoly~B. Kolomeisky.
\newblock Simple mechanochemistry describes the dynamics of kinesin molecules.
\newblock {\em Proceedings of the National Academy of Sciences of the United
  States of America}, 98(14):7748--7753, 2001.

\bibitem{Derrida1997}
B.~Derrida, J.~L. Lebowitz, and E.~R. Speer.
\newblock Shock profiles for the asymmetric simple exclusion process in one
  dimension.
\newblock {\em Journal of Statistical Physics}, 89:135--167, 1997.

\bibitem{Qian1997}
Hong Qian.
\newblock A simple theory of motor protein kinetics and energetics.
\newblock {\em Biophysical Chemistry}, 67:263--267, 1997.

\bibitem{Gross2002}
Steven~P. Gross, Michael~A. Welte, Steven~M. Block, and Eric~F.
Wieschaus.
\newblock Coordination of opposite-polarity microtubule motors.
\newblock {\em The Journal of Cell Biology}, 156:715--724, 2002.

\bibitem{Deacon2003}
Sean~W. Deacon, Anna~S. Serpinskaya, Patricia~S. Vaughan,
Monica~Lopez
  Fanarraga, Isabelle Vernos, Kevin~T. Vaughan, and Vladimir~I. Gelfand.
\newblock Dynactin is required for bidirectional organelle transport.
\newblock {\em The Journal of Cell Biology}, 160:297--301, 2003.

\bibitem{Welte2005}
Michael~A. Welte, Silvia Cermelli, John Griner, Arturo Viera,
Yi~Guo, Dae-Hwan
  Kim, Joseph~G. Gindhart, and Steven~P. Gross.
\newblock Regulation of lipid-droplet transport by the perilipin homolog lsd2.
\newblock {\em Current Biology}, 15:1266--1275, 2005.

\bibitem{Smith2004}
G.~A. Smith, L.~Pomeranz S.~P. Gross, and L.~W. Enquist.
\newblock Local modulation of plus-end transport targets herpesvirus entry and
  egress in sensory axons.
\newblock {\em Proceedings of the National Academy of Sciences of the United
  States of America}, 101(45):16034--16039, 2004.

\bibitem{Lipowsky2008}
Melanie~J.I. M\"{u}ller, Stefan Klumpp, and Reinhard Lipowsky.
\newblock Tug-of-war as a cooperative mechanism for bidirectional cargo
  transport by molecular motors.
\newblock {\em Proceedings of the National Academy of Sciences of the United
  States of America}, 105(12):4609--4614, 2008.

\bibitem{Beeg2008}
Janina Beeg, Stefan Klumpp, Rumiana Dimova, Rub\`{e}n~Serral
Graci\`{a},
  Eberhard Unger, , and Reinhard Lipowsky.
\newblock Transport of beads by several kinesin motors.
\newblock {\em Biophysical Journal}, 94:532--541, 2008.

\bibitem{Lipowsky2005}
Stefan Klumpp and Reinhard Lipowsky.
\newblock Cooperative cargo transport by several molecular motors.
\newblock {\em Proceedings of the National Academy of Sciences of the United
  States of America}, 102:17284--17289, 2005.

\bibitem{Lipowsky20081}
Melanie~J.I. M\"{u}ller, Janina Beeg, Rumiana Dimova, Stefan Klumpp,
and
  Reinhard Lipowsky.
\newblock Traffic by small teams of molecular motors.
\newblock {\em arXiv:0807.0964v1 [cond-mat.stat-mech]}.

\bibitem{Lipowsky20082}
Melanie~J.I. M\"{u}ller, Stefan Klumpp, and Reinhard Lipowsky.
\newblock Motility states of molecular motors engaged in a stochastic
  tug-of-war.
\newblock {\em Journal of Statistical Physics}, 133:1059--1081, 2008.

\bibitem{Chen2002}
Yi~der Chen, Bo~Yan, and Robert~J. Rubin.
\newblock Fluctuations and randomness of movement of the bead powered by a
  single kinesin molecule in a force-clamped motility assay: Monte carlo
  simulations.
\newblock {\em Biophysical Journal}, 83:2360--2369, 2002.

\bibitem{Elston2000}
Timothy~C. Elston and Charles~S. Peskin.
\newblock The role of protein flexibility in molecular motor function: Coupled
  diffusion in a tilted periodic potential.
\newblock {\em SIAM Journal on Applied Mathematics}, 60(3):842--867, 2000.

\bibitem{Frank1995}
Frank J\"{u}licher and Jacques Prost.
\newblock Cooperative molecular motors.
\newblock {\em Physical Review Letters}, 75:2618--2621, 1995.

\bibitem{Block1994}
K.~Svoboda and S.M. Block.
\newblock Force and velocity measured for single kinesin molecules.
\newblock {\em Cell}, 77:773--784, 1994.

\end{thebibliography}
\end{document}